\documentclass[12pt,a4paper]{article}
\usepackage{epsfig}
\def\lt{\raisebox{0.2ex}{$<$}}
\def\gt{\raisebox{0.2ex}{$>$}}

\title{Light-particle emission from the fissioning nuclei\\ $^{126}$Ba, 
        $^{188}$Pt and $^{266,272,278}$110:\\ theoretical predictions and 
        experimental results
\thanks{The work is partly supported by the Polish
Committee of Scientific Research under Contract No. 2P03B~011~12}
}
\author{
  K.~Pomorski$^a$, B.~Nerlo-Pomorska$^a$, A.~Surowiec$^a$, M.~Kowal$^a$,\\
  J.~Bartel$^b$, K.~Dietrich$^c$, J.~Richert$^d$, C. Schmitt$^b$,\\ 
  B.~Benoit$^{b,e}$, E. de Goes Brennand$^e$\footnote{On leave from 
  Universidade Estadual da Paraiba, Brasil}, L.~Donadille$^f$, C.~Badimon$^g$\\
{\it $^a$Katedra Fizyki Teoretycznej, Uniwersytet Marii Curie-Sk\l odowskiej,}\\
{\it Lublin, Poland}\\
{\it $^b$Institut de Recherches Subatomiques, Strasbourg, France}\\
{\it $^c$Physik Department, Technische Universi\"at M\"unchen, Germany} \\
{\it $^d$Laboratoire de Physique Th\'eorique,  Universit\'e Louis Pasteur,} \\ 
{\it Strasbourg, France}\\
{\it $^e$Universit\'e Libre de Bruxelles, Belgium}\\
{\it $^f$University of Birmingham, United Kingdom}\\  
{\it $^g$Centre d'Etudes Nucl\'eaires de  Bordeaux-Gradignan, France}
}

\date{}

\begin{document}

\maketitle

\noindent
\begin{abstract}
We present a comparison of our model treating fission dynamics in
conjunction 
with light-particle ($n, p, \alpha$) evaporation with the available 
experimental data for the nuclei $^{126}$Ba, $^{188}$Pt and three isotopes 
of the element Z=110. The dynamics of the symmetric fission process is described 
through the solution of a classical Langevin equation for a single collective 
variable characterizing the nuclear deformation along the fission path. 
A microscopic approach is used to evaluate the emission rates for pre-fission 
light particles. Entrance-channel effects are taken into account by 
generating an initial spin distribution of the compound nucleus formed by
the fusion of two deformed nuclei with different relative orientations.
\end{abstract} 

                                                             \\[2.0ex]
  

\newpage
\section{Introduction}

The dynamical time evolution of the fission process from an initially formed 
compound nucleus with a more or less compact shape to the saddle and scission 
configurations and the simultaneous emission during this deformation process 
of light particles constitutes a very complex problem. 
It is the aim of our study to describe this process taking into account 
the excitation of both thermal and rotational nature, the initial configuration  
and to calculate the evaporation of light particles from an excited, 
rotating, deformed nucleus on its way from the initial compact shape to the 
scission point.

In the absence of a complete microscopic ab initio theory of such a dynamical 
process, many different theoretical approaches aiming at its description have 
been worked out \cite{SDP91}-\cite{Prz94}. They generally
rely on a classical description of the evolution of the 
collective coordinates which are introduced as abundant variables. 
In these descriptions, collective parameters appear 
(collective mass, friction, and diffusion coefficients) which depend on the 
collective coordinates. It seems clear that the quality of the theoretical 
description will depend on the more or less pertinent choice of the collective 
coordinates and the degree of realism of the underlying theory used to 
determine the collective parameter functions. 
The importance of pure quantum effects such as pairing correlations and
the existence of shell effects  which are present at low excitation energies 
has been investigated recently \cite{Hof97}-\cite{IH99}.

The multiplicity and the characteristics (energies, angular distributions, 
correlation functions) of prefission light particles and gamma emission may 
provide some useful information on the time evolution of the nucleus as 
it evolves towards the saddle and scission point. Observables like prefission 
light-particle multiplicities may, indeed, work as a clock. Angular 
distributions and particle correlations may give informations on the
surface and deformation of the fissioning nucleus.

We  recently developed a realistic time dependent model which described 
the time evolution of an excited nucleus as it is e.g. generated in a
heavy-ion collision and decaying through symmetric fission with pre- and
post-fission 
light particle emission \cite{PBR96}. We  applied our description to the 
case of $^{160}$Yb and made a detailed analysis of different physical 
quantities characterizing the decaying system. We found a rather good
agreement of our theoretical predictions with the experimentally observed
light-particle multiplicities.

More recently, several new sets of experimental data have become available 
\cite{Bad98}-\cite{Kal99} with a more complete analysis of the emitted light 
particles. It is the aim of the present work to systematically test our
approach by a comparison with these experimental data. 
The fact that these data are obtained in quite different 
regions of the periodic table makes a comparison between calculated and
experimental results all  the more challenging.

A particular  motivation for this work is the explicit description of the 
entrance channel through which the initial spin distribution of the system
is generated. This is achieved by considering all possible relative orientations 
of the two colliding deformed nuclei in the entrance channel. Their different 
orientations and the different possible impact parameters of the reaction 
yield the initial spin distribution $d\sigma/dL$ versus $L$ which
determines the relative weights of the initial angular momentum of the compound
nucleus with which the Langevin trajectories are started.

Still another motivation for the present work is the presentation of 
the link between the well known Weisskopf formula \cite{We37} and a new 
and more microscopically founded description of the emission width for the 
light particles in terms of phase-space densities of the particle to be 
emitted \cite{DPR95}.

The paper is organized as follows: The dynamical model describing the time 
evolution of a deformed, hot, and rotating nucleus from its initial to a
final configuration is presented in section 2. In section 3 we specify 
how to determine the emission rates for light particles from a hot,
deformed and rotating nucleus moving along the previously described trajectory. In
section 4 we show how we obtain the initial spin distribution which has been not 
considered in our previous study \cite{PBR96}. Our results concerning the 
nuclear systems $^{126}$Ba, $^{188}$Pt and $^{266,272,278}$110 are presented in 
section 5. The paper is closed in section 6 with a summary of the most
relevant results and an outlook on planned extensions of the model. 
\vspace{1cm}
\section{Collective dynamics of an excited system }

We consider an ensemble of deformed nuclei with finite  excitation
energies and 
rotational angular momenta as given by the initial conditions determined 
from the entrance channel. The subsequent time evolution of the nucleus is 
governed in our present description by a single collective coordinate 
$q = R_{12}/R_0$ where $R_{12}$ is the distance between the two centers of 
mass of the left-right symmetric deformed nucleus and $R_0$ is the radius
of the corresponding spherical nucleus having the same volume. This collective 
variable is defined in the framework of a Trentalange--Koonin--Sierk (TKS) 
\cite{TKS80}
parameterization of the surface of the nucleus. The TKS deformation parameters 
are related to $q$ by means of a minimization procedure of the collective
potential energy  defined below \cite{BMR96}.

Denoting  the conjugate momentum by $p(t)$ we use the following classical 
equations of motion to describe the time evolution of the fissioning nucleus
\cite{PBR96}
\begin{eqnarray}
  &&  {dq\over dt} = {p \over M(q)}  \label{c2.1}\\
  && {dp\over dt} = {1\over 2} \left({p\over M(q)}\right)^2\,
     {dM(q)\over dq} - {dV(q)\over dq} - {\gamma(q)\over M(q)}p + F_L(t)
\, .
\label{c2.2}
\end{eqnarray}
Here $M(q)$ is the collective mass determined in the incompressible fluid 
approximation \cite{Da76} and $\gamma(q)$ the friction coefficient calculated 
in the wall-and-window friction model \cite{BRS77}. The collective potential 
V(q) is obtained as the difference of the Helmholtz free energy at deformation 
$q$ minus the one for the ground-state deformation. 

The free energy could in principle be obtained from a microscopic mean-field 
calculation at finite temperature. 
To perform such a constraint Hartree-Fock calculation using a reasonable 
effective nucleon-nucleon interaction of the Skyrme or Gogny type at every 
point in the multidimensional deformation space is, of course, completely out 
of question due to the tremendous computer time such an analysis would involve. 
Even to perform the same kind of calculation on the level of a selfconsistent 
semiclassical approximation like the Extended Thomas-Fermi (ETF) method \cite{xx1}
at finite temperature \cite{xx2}, which would describe the average nuclear 
structure without shell oscillations, would be far too time consuming. This is
why we have rather used a still simpler semiclassical approach and the liquid drop 
model with the parametrization of Myers and Swiatecki \cite{MS67} as explained in
Appendix A. We have tested the barrier heights obtained in this way versus 
the ones obtained from selfconsistent semiclassical calculations (at zero 
temperature) and obtained agreement within a few MeV as described in 
Ref.\cite{BMR96}.
These calculations describe, of course, deformation properties of non excited 
nuclei. In order to take nuclear excitation into account we have used the 
temperature 
dependence of the LDM parameters associated with the Skyrme SkM$^*$ interaction 
\cite{xx3}, determined for that interaction through selfconsistent semiclassical 
calculations at finite temperature in Ref. \cite{BMR96}. 

To use the temperature dependence 
of the LDM parameters associated with the Skyrme SkM$^*$ force together with 
the Myers-Swiatecki parametrization of the LDM is, of course, inconsistent. 
Since approaches give very similar results for the semiclassical energy 
at zero temperature this approximation seems, however, very reasonable.
The semiclassical approach used here is of course only an approximation at 
low temperatures since shell effects are absent from our description. It is 
however well known that nuclear shell effects are washed out with increasing 
nuclear temperature and have essentially disappeared beyond 
T$\simeq$ 2.5-3 MeV. At these temperatures the semiclasssical results are 
becoming exact. For the cold systems 
the fission barriers obtained with this potential are higher that those evaluated 
by Sierk \cite{Sie86}, but already at excitation energies $E^*\approx 90$ MeV 
both barriers become comparable and for $E^* \gt 90$ MeV the barriers evaluated 
in our model are even smaller than the Yukawa folded ones of Sierk.

The friction term and the Langevin force $F_L(t)$ in Eq. 
(\ref{c2.2}) generate  the irreversible production of heat energy and the energy 
fluctuations respectively which both originate from the coupling of the 
collective dynamics to the intrinsic degrees of freedom. In practice one defines 
$F_L(t) = \sqrt{D(q)} f_l(t)$, where $D(q)$ is the diffusion coefficient.
We take the simplifying point of view that it is related to the friction 
coefficient $\gamma(q)$ through the Einstein relation $D(q) = \gamma(q) \,T$, 
where $T$ is the temperature of the system. The quantity $f_L(\tau)$ can 
be written in the form $f_L(\tau) = \sqrt{\tau}\,\eta$, where $\tau$ is
a time step length corresponding to a time interval $[t, t + \tau]$ and
$\eta$ is a gaussian distributed random number with zero average 
$\langle\eta\rangle \!=\! 0$ and variance $\langle\eta^2\rangle \!=\! 2$ 
where brackets represent ensemble averages.

The Einstein relation is in principle valid at high temperature, in a 
regime where the process can be described in a classical framework.
Quantum effects (pairing, shell corrections, collective shape vibrations) 
may be present at low
temperatures and modify the relation between $D$ and $\gamma$ 
\cite{Yam98}-\cite{IH99}. We shall come back to this point later on.

In principle, one would have to treat each ensemble of nuclei with
a given initial angular momentum $L$ and a given initial excitation
energy microcanonically. We assume that we may replace this 
microcanonical ensemble by a grand canonical one in which, instead of
the mean excitation energy, the temperature $T$ has a well-defined value
at a given time. This simplifying assumption is innocuous as far as 
the \underline{mean values} of observables are concerned, but it may 
falsify the fluctuations to a certain extent. 
We assume that the time scale 
which governs the fission dynamics is much larger than the internal 
equilibration time, otherwise the definition of a nuclear temperature would 
make no sense. Under these conditions the system can be considered as being 
continuously close to equilibrium. We suppose that the nuclear excitation 
energy $E^*$ is related to $T$ through the usual Fermi-gas expressions 
$E^* = a(q)T^2$, where $a(q)$ is the level density parameter at a given 
deformation $q$.

For practical calculations Eqs. (\ref{c2.1}) and (\ref{c2.2}) are rewritten 
in a discretized form and numerically integrated over small time steps $\tau$ 
\cite{PBR96}. In order to perform such an integration one needs to start 
from some fixed initial conditions which define the beginning of the process.
We take into account the possible emission of light particles in every time 
step and we make sure that the average total energy of the system will be
conserved.
This determines the nuclear excitation energy and hence the temperature at 
each instant of time.
                                                                    
\section{Particle emission from a hot, deformed, and rotating
          nucleus} 

Particle emission before scission is governed by transition rates
$\Gamma^{\alpha\beta}_{\nu}(E^*, L)$ which determine the number of particles
of type $\nu$ (we take into account neutrons, protons, and $\alpha$
particles) emitted per unit time with an energy $e_\alpha $ in an interval
$\left[e_\alpha - {1 \over 2} \Delta e, e_\alpha + {1 \over 2} \Delta e\right]$ 
and with an angular momentum $\ell_\beta$ from a nucleus with average excitation 
energy $E^*$ and total angular momentum $L$. In Ref. \cite{PBR96} we used 
the well known Weisskopf formula \cite{We37} for the partial width 
$\Gamma^{\alpha\beta}_{\nu}(E^*, L)$ in terms of densities of states of
the emitting and residual nucleus and of the transmission coefficient 
$w_\nu(e, \ell_\beta)$ for emitted a particle $\nu$ with given energy $e$
and angular momentum $\ell_\beta$. The determination of $w_\nu$ takes into 
account the deformation and rotation of the emitting nucleus (see Appendix
B of Ref. \cite{PBR96} for details).

In Ref. \cite{DPR95} another more microscopic  
determination of these transition rates was proposed. In this 
framework the transition rates $\Gamma^{\alpha\beta}_\nu$ are given as 
\begin{equation}
  \Gamma^{\alpha\beta}_\nu = {d^2 n_\nu \over d\varepsilon_\alpha
d\ell_\beta}  
                             \Delta\varepsilon \,\Delta \ell \, ,
\label{c3.1}
\end{equation}
where $\varepsilon_\alpha$ and $\ell_\beta$ characterize an emission
energy and angular momentum lying in the intervals
$$
   \left[\varepsilon_\alpha - {1\over 2}\Delta\varepsilon,\,
      \varepsilon_\alpha + {1\over 2}\Delta \varepsilon\right] \;\;\;\; 
      {\rm and} \;\;\;\;
  \left[\ell_\beta - {1\over 2} \Delta \ell \, , \, \ell_\beta + {1\over 2} 
      \Delta \ell \right] 
$$
respectively.

The number $n_\nu$ of particles of type $\nu$ which are emitted per time
unit through the surface $\Sigma$ of the fissioning nucleus is given by
\cite{DPR95}
\begin{equation}
 n_\nu = \int\limits_\Sigma \! d\sigma \int \! d^3 p' \, f_\nu(\vec r_0', 
 \vec p\,') \, v'_\perp(\vec r_0') \, w_\nu( v'_\perp (\vec r_0'))
\label{c3.2}
\end{equation}
where $\vec p\,'$, $\vec v\,'$ are the momentum and velocity in the
body-fixed frame. The quantity $v'_\perp$ is the velocity component 
perpendicular to the emission surface at the surface point $\vec r_0'$. The 
$\vec p\,' = m \vec v\,' + m \vec\omega \!\times\! \vec r\,'$ is the momentum
of the particle of mass $m$ in the laboratory reference frame and $\vec\omega$ 
the angular velocity of the nucleus in this frame. Here and henceforward, primed 
quantities refer to the body-fixed frame.

The classical distribution in phase space reads
\begin{equation}
   f_\nu(\vec r\,' , \vec p\,') = {2 \over h^3}
   {\theta(\vec r\,') \over 1 + \exp \left[ {1 \over T}
  \left({p'^2 \over 2m} + U - \omega \ell' - \mu_\nu \right)\right] } \,.
\label{c3.3}\end{equation}
The $\theta$ function is 1 if $\vec r\,'$ lies inside the nuclear volume 
or on its surface $\Sigma$ and zero otherwise. 
The quantity $\mu_\nu$ is the chemical potential and $\ell'$ the
body-fixed angular momentum in the direction perpendicular to the axis of
rotational symmetry of the deformed nucleus. The potential $U$ is taken as
\begin{equation}
   U(\vec r\,') = -V_0 + V_{Cb} (\vec r\,') \,,
\label{c3.4}\end{equation}
where $V_0 \!>\! 0$ is chosen as a constant mean field potential and
$V_{\rm Cb}$ is the Coulomb potential experienced at $\vec r\,'$ by
protons.

The quantity $w_\nu(v'_\perp (\vec r_0'))$ is the classical
transmission coefficient for the emission of a particle of type $\nu$.
The transmission factor $w_\nu$  was chosen to be the one of an inverted
harmonic oscillator \cite{PBR96}.
    
The explicit relation of (3) with the Weisskopf formulation is discussed in
Appendix B for a spherical  emitting nucleus with angular momentum $L=0$. 
The rates given by (3) can be worked out numerically if the distribution
function in the phase space is known, what is the case for the neutrons
and protons but not for the $\alpha$-particles. We are working on a model
which describes the distribution of $\alpha$-particles by folding of the
distributions of four (2$n$ and 2$p$) correlated particles.
The present approach is like the one of Ref. \cite{PBR96} and it allows 
us, in principle, to determine the particle emission in a given direction 
of space, hence the determination of observables
like angular distributions and particle-particle correlation functions.
Such observables may be worked out in the future. We did not attempt 
to do it here because of the lack of corresponding experimental data.

The transition rates $\Gamma^{\alpha\beta}_\nu$ are used in a simulation
algorithm by means of which we determine at each time step $[t,t+\tau]$
along each classical trajectory whether a particle of given type with
an energy and angular momentum in given intervals is emitted or not from
the compound nucleus. Since the algorithm is already described in detail
in Ref. \cite{PBR96} we do not repeat it here.

\subsection{Initial conditions and energy balance}

In order to integrate the classical equations (1) and (2) we need to fix
the initial conditions from which the compound system starts and
evolves either through a fission channel to its saddle and scission
point or stays as a compound system which only emits light particles, i.e.
ends up as an evaporation residue.

All the experimental systems which are considered in the next section
are generated by means of heavy ion collisions at some bombarding
energy. The nuclei which are involved in the reaction process can be
deformed.  The initial conditions corresponding to the origin of time
are fixed by $q_0$ and $p_0$, the initial value of the collective
variable and its conjugate momentum and the spin distribution of the
system which fixes the relative weight of the angular momentum of the
initial compound systems.

The coordinate $q_0$ is fixed at the value of $q$ where the collective
potential $V(q)$ is minimal and its conjugate momentum is drawn from a
normalized gaussian distribution
\begin{equation}
 P(p_0) = (2\pi M T_0)^{1/2} \exp(-p^2_0/2MT_0)\,,
 \end{equation}
where $M = M(q_0)$ is the collective mass. The initial temperature
$T_0$ is obtained through $E^*_0 = a(q_0)T^2_0$, where $E^*_0$ is the
initial excitation energy which can be obtained from the knowledge of the
total energy as explained below. 

In the reaction process, the compound nucleus can be formed with
different values of the angular momentum.  If both nuclei are spherical
it is easy to construct the initial spin distribution under the
assumption that the reaction cross section is given by the geometrical
expression
\begin{equation}
 {d\sigma(L)\over dL} = {\pi\over k^2}(2L + 1)\,,
\end{equation}
where $k$ is the wavenumber of the relative motion.

The situation is more complicated if one or both initial nuclei are
deformed. In
principle one should then consider all possible relative orientations 
of the nuclei and follow their relative trajectories from an infinite
distance up to fusion. This implies the knowledge of the
nuclear and Coulomb potential for given relative distance between the
centers of mass and given orientation. This raises intricate
problems which have been considered by several authors
\cite{AH76,TRI99,PPR94}
but not yet solved in their full generality.

In a first step we want to avoid such cumbersome calculations and
restrict ourselves to a simplified procedure which has been already proposed
in former work \cite{PPR94,PPB99}. In practice we consider the scattering
of spherical ions of fixed energy $E$ and introduce the gaussian energy
distribution with the width 
\begin{equation} 
\Delta E =\delta E + \Delta_1 E + \Delta_2 E\,, 
\end{equation} 
where $\delta E$ is the experimental beam width, $\Delta_1 E$ describes 
the slowing down of the projectile in the target, and the  quantity
$\Delta_2E$ 
is the difference of the Coulomb barriers
between two extreme geometries of touching spheroidally  deformed nuclei,
i.e. with aligned (tip on tip) or parallel (side on side) $z$--axes.

One can now work out the classical trajectories describing the relative
motion  of the equivalent spherical ions for every initial angular
momentum $L$ in
$[0,L_{\rm max}]$, where $L_{\rm max}$ is the maximal angular momentum
for the colliding ions and for a sufficient number of energy values within
the distribution defined above.

Repeating the trajectory calculations leads to the spin distribution
\begin{equation}
  \left({d\sigma_F\over dL}\right)_{L_i} = {2\pi\over k^2}L_i \,
     {N^F_i\over N_i}\,,
     \end{equation}
where $L_i$ is the considered angular momentum, $N^F_i$ is the number
of trajectories which lead to fusion and $N_i$ is the total number of
trajectories. The quantity $k$ is the wave--number of
 the relative motion of the incident nuclei.  The present procedure, 
even though it is not rigorous,
leads to spin distributions which are rather realistic \cite{PPR94,PPB99}.

Finally, in order to describe the evolution of the excited compound
system one needs to follow the evolution of the average excitation energy 
 along the trajectory. 
This is achieved by the requirement of energy conservation.

At the initial point of a trajectory the
total available energy can be written as 
\begin{eqnarray} 
 E_{\rm tot} &=& E_{\rm coll} + E_{\rm rot} + E^*_0 \nonumber  \\
 &=& {p^2_0\over 2M(q_0)} + V(q_0) + {L^2_0\over 2J(q_0)} + E^*_0\,.
\end{eqnarray} 
Here, $J(q_0)$ is the moment of inertia of the compound system
taken as a rigid deformed rotator and $E^*_0$ is the initial intrinsic
excitation energy of the compound nucleus. The intrinsic excitation energy 
$E^*(t)$ at any given later time $t$ can 
be determined from the energy balance 
\begin{equation} 
 E_{\rm tot} = E_{\rm coll} + E_{\rm rot} + E^*_0 + B_\nu + e_\alpha + 
 E_{\rm recoil}\,,
\end{equation}
where $B_\nu$ is the binding energy of the emitted particle, different
from zero only for $\alpha$-particles, $e_\alpha$ is the kinetic energy
of the emitted particle and $E_{\rm recoil}$ is the recoil energy of
the nucleus after the emission of a particle, which can be neglected 
in practical calculations.

For each choice of the initial conditions one generates a separate
trajectory. Emitted particles are counted with their energy and angular
momentum. If the system overcomes the fission barrier the trajectory
(event) contributes to the final fission cross section 
\begin{equation}
 \sigma_{\rm fiss} = \sum_i {d \sigma_{\rm fiss} \over d L_i} =
 \sum_i \left( {d \sigma_F \over d L_i}\right)_{L_i}
 \cdot { N_i^{\rm fiss} \over N_i^F } \,, 
\end{equation}
where $N_i^{\rm fiss}$ is the number of trajectories which lead to fission,
and $N_i^F$ -- the number of fused trajectories at a given angular momentum
$L_i$. The sum runs over all angular momentum bins and $d\sigma_F/dL$ is
the fusion cross section given by Eq. (10).

The numbers of prefission particles obtained for each angular momentum of
the compound nucleus ($M_\nu$) are weighed with the differential fission 
cross section in order to obtain the measured number of particles emitted 
in coincidence with fission:
\begin{equation}
\lt M_\nu \gt = {\sum_i {d \sigma_{\rm fiss} \over d L_i} \cdot 
                 M_\nu (L_i) \over    \sigma_{\rm fiss}} \,.
\end{equation}

\section{Application of the model to various experimentally studied systems}

In the present section we aim to confront experimental results
concerning the multiplicities of emitted particles in coincidence with
symmetric fission events with the model described in the preceding section.
All measured and calculated data are presented in Table 1 for three systems:
$^{126}$Ba, $^{188}$Pt, and $^{266,272,278}$110 obtained by different entrance
channels and at different energies. These data result from experiments 
performed at the SARA (Grenoble) and the VIVITRON (Strasbourg) 
\cite{Bad98}-\cite{BD99} using the DEMON neutron detector 
\cite{Mos94,Til95}. Unfortunately, the charged particles ($p, \alpha$)
were not measured, so we give in Table 1 the theoretical
predictions only. We present a more detailed discussion of the results for
each system in the next subsections. 

\subsection{The $^{126}$Ba compound system}

The compound nucleus $^{126}$Ba has been experimentally produced 
through two different entrance channels:  

\medskip
\noindent
(I)~~~  $^{28}{\rm Si} + ^{98}{\rm Mo} \rightarrow ^{126}{\rm Ba}$
        ~~~~at~ $E_{lab}$=204, 187, and 166~MeV\\
and  

\medskip
\noindent
(II)~~~ $^{19}{\rm F} + ^{107}{\rm Ag} \rightarrow ^{126}{\rm Ba}$
        ~~~~at~ $E_{lab}$=148 and 128~MeV. 

\medskip
\noindent
Fig. 1 shows the fusion cross-section and the fission yields as a 
function of the initial angular 
momentum. The distributions of the initial angular momentum
were calculated with the help of the corresponding Langevin equation
\cite{Fro90,Prz94,PPR94} for two different entrance channels (I, II) 
and various bombarding energies. As it can be seen, the fission yields 
are rather small and located in the tails of the distributions. 
This observation which also holds for many other systems implies that
it is of great importance to calculate carefully the dependence of the
fusion cross-section on the angular momentum if one wants to describe
the competition between the decay of the compound nucleus by fission
and by light particle emission in a correct way. Concerning Fig.~1 we
also note that the dominant part of the initial excitation energy
resides in the collective rotational motion. 
One observes in Fig. 1 a general increase of the fission cross section 
with the excitation energy of the compound nucleus $^{126}$Ba with  
the only exception of the highest excitation energy. In that case the 
emission of $\alpha$-particles becomes important and competes with 
the fission process. This causes a substantial loss of internal energy 
due to particle emission which in turn leads to an increase of the fission 
barrier and therefore to a smaller fission yield. 

In Fig.~2, the potential energy $V_{\rm fiss}$ of the fissioning nucleus 
$^{126}$Ba for different angular momenta is plotted as a function of the
relative distance between fission fragments. One can see in Fig.~2
that the fission barrier of the fused nucleus at temperature $T$=1.6 MeV 
and angular $L$=0 is quite high in our model. It is equal  to 43~MeV.
It vanishes at very large angular momentum ($L \approx 80\hbar$). 
The high temperature can also reduce significantly the fission barrier
as one can learn from Fig.~3, where the free energy of $^{126}$Ba is plotted 
for a few temperatures between 0 and 4.8 MeV. 
This is why one can observe a significant fission
rate only at large angular momenta and  high excitation energies.
All curves representing the fission barriers in Figs. 2 and 3 end at the 
scission point region. It is seen from Figs. 2 and 3 that the saddle
and scission points are close to each other for $^{126}$Ba.
 
Part of the excitation energy of $^{126}$Ba is contained in the rotational
mode. In Fig.~4, the temperature of $^{126}$Ba corresponding to different 
excitation energies is plotted as a function of angular momentum. One can 
see that, at the lowest excitation energy $E^*=$84 MeV and the highest 
angular momentum, the temperature is 0.5~MeV only. This means that our 
theoretical predictions based on the statistical equilibrium could be too 
rough in this case.

The experimental neutron multiplicities obtained for both the fission
and the fusion channel are shown in Table 1 along with the calculated neutron, 
proton, and $\alpha$-particle multiplicities.

In Figs.~5 and~6, we show the calculated average numbers of
neutrons, protons, and $\alpha$-particles emitted in coincidence
with fission as a function of the excitation energy $E^*$ 
of the initial compound nucleus. The initial compound nucleus $^{126}$Ba 
is produced by the fusion reaction (I) in Fig.~5, and in Fig.~6 the same 
initial compound nucleus originates from the fusion reaction (II).
Only the $n$-emission on the way to fission has been measured so
far. The measured numbers of prefission neutrons are indicated
by points with error bars.
At two excitation energies we have measurements for the two
different entrance channels. As one can read from Table 1 at the excitation 
energy $E^*$ = 101.4~MeV the measured and calculated numbers of
emitted prefission neutrons (in coincidence with fission) are
$M_n$ = 1.32 and 1.83, resp., for the entrance channel
(I) (Fig.~5) and $M_n$ = 1.31 and 1.80, resp., for the entrance
channel II (Fig.~6). At the higher excitation energy $E^*$ = 118.5~MeV 
the measured and calculated numbers of emitted
neutrons are $M_n = 2.01$ and 1.71 for the entrance channel (I)
(Fig.~5) and $M_n = 1.85$ and 1.99 for the entrance channel (II)
(Fig.~6).

The experimentally observed number of emitted prefission
neutrons at given excitation energy $E^*$ of the
initial compound nucleus is thus larger for the entrance channel
(I) than for the entrance channel (II) and the difference of the
number of emitted neutrons is seen to be larger for the larger
excitation energy.

The interpretation of this observation is that for
$^{126}$Ba only the compound nuclei formed with high
angular momentum may undergo fission and thus give rise to
prefission neutrons because the fission barrier for low angular
momentum is high ($\geq$ 15~MeV). For $T$=1.6 MeV the LD fission
barriers vanish for about $L = 70~\hbar$
as can be seen from Fig. 2. Thus the main part of the prefission
neutrons is obtained for angular momenta around $L = 80~\hbar$
as indicated in Fig. 1. At given excitation energy $E^*$
the number of compound nuclei formed with such high angular
momentum is larger for the entrance channel (I) with
$^{28}$Si as a projectile than for the entrance
channel (II) with $^{19}$F as a projectile (see Fig. 1).

The theoretically estimated prefission proton multiplicities ($M_p$)
are very small as one can see in Figs. 5 and 6. The multiplicities
of $\alpha$-particles are not negligible at the highest ($E^*$=131.7 MeV)
and the lowest ($E^*$=84.1 MeV) excitation energies (see Figs. 5 and 6).
The increase of $M_\alpha$ at high excitation energy is mostly due to 
the high temperature effects while the unexpectedly large number of emitted 
$\alpha$-particles at $E^*$=84.1 MeV is mostly due to the large deformation 
of the emitter and the huge centrifugal potential ($L \ge 80 \hbar$) 
reducing the  already small Coulomb barrier for the $\alpha$-particles 
at the tips even further.

As for the comparison of measured and calculated numbers of
emitted prefission neutrons, the general agreement is not
unsatisfactory. Nevertheless, one notices that the calculated
average emission numbers are larger than the observed ones for
the higher excitation energy $E^*$ = 118,5~MeV. If,
instead of the temperature-independent friction, which underlies
these calculations, we would use a friction parameter increasing
with temperature, the agreement might be better. 

\subsection{The $^{188}$Pt compound system}
 
As seen in Table~1, the compound nucleus $^{188}$Pt was produced
at two excitation energies $E^*$=100~MeV and 66.5~MeV using two different
reactions \cite{Bad98}
\begin{itemize}
\item $^{34}{\rm S}~+~^{154}{\rm Sm} \rightarrow ^{188}{\rm Pt}$
        ~~~~at~ $E_{lab}$= 203 and 160 MeV ,
 
\item $^{16}{\rm O}~+~^{172}{\rm Yb} \rightarrow ^{188}{\rm Pt}$
        ~~~~at~ $E_{lab}$= 138 MeV. 
\end{itemize}
The theoretical estimates of the initial spin distributions (solid lines)
as well as the fission rates (bars) for $E^*$=100 MeV and two entrance 
channels as well as for $E^*$=66.5 MeV are plotted in Fig.~7 as functions
of angular momentum of the compound nucleus.
It is seen that each reaction leads to different spin
distribution. These differences are responsible for the entrance channel 
effects in the prefission neutron multiplicities emitted by $^{188}$Pt
at $E^*$=100 MeV (see Table 1).

Similarly, as in  the case of $^{126}$Ba, we present in Fig.~8 the fission
barriers of $^{188}$Pt obtained for different angular momenta ($L$) at fixed
temperature $T$=1.6~MeV. 
It is seen that for $T=1.6$~MeV and $L=70\,\hbar$ 
the fission barrier becomes negligible. Also one can see in Fig.~9 that with 
increasing temperature the fission barrier decreases. At $T=4.8$~MeV and 
$L=30\,\hbar$, the fission barrier of $^{188}$Pt is about four times lower than
for $T=0$ and $L=30\,\hbar$. 
Contrary to the case of $^{126}$Ba, where the saddle point is very close to 
the scission point, the path from the saddle point ($q=R_{12}/R_0 \approx$ 1.6)
to the scission point ($q \approx$ 2.2) of $^{188}$Pt is much longer 
and will take more time. This implies that the case of $^{188}$Pt 
is better suited to study the influence of dynamical effects on the prefission 
light-particle multiplicities.

The prescission neutron multiplicities theoretically predicted for
$^{188}$Pt are too small (by $\sim$1 unit on the average) in comparison 
with the experimental data \cite{Bad98}.
Also the effect of the entrance channel is not fully reproduced.
Unfortunately the protons and $\alpha$-particle multiplicities were not 
measured, so we cannot test the predictive power of our model in this case.

\subsection{The $Z=110$, $N=156,162,$ and $168$ compound systems}

Three isotopes of the superheavy compound nucleus with $Z=110$ were formed by 
means of the following fusion reactions
\begin{itemize}
\item $^{58}$Ni + $^{208}$Pb $\rightarrow$ $^{266}$110  \cite{Don98}
\item $^{64}$Ni + $^{208}$Pb $\rightarrow$ $^{272}$110  \cite{Don98}
\item $^{40}$Ca + $^{232}$Th $\rightarrow$ $^{272}$110  \cite{BD99}
\item $^{40}$Ar + $^{238}$U  $\rightarrow$ $^{278}$110  \cite{BD99}
\end{itemize}
at different excitation energies between 66 MeV and 186 MeV. 
The prefission neutron multiplicities were determined experimentally
and estimated theoretically within our model. Both the experimental and
theoretical results are given in Table 1. 

Using  Langevin type equations to describe the fusion process 
\cite{Fro90,Prz94,PPR94}  we obtained the
initial spin distribution of compound nucleus. As an example, the differential 
fusion cross section is plotted in Fig.~10 for the reaction $^{40}$Ca+
$^{232}$Th at $E_{lab}$=250.4 MeV
as a function of the angular momentum of the system. The effective fission 
cross section is marked by bars in the figure. It is seen in Fig.~10 that
only the lowest angular momenta contribute to the fusion fission process
of $^{272}$110 due to the fact that the fission barrier of $^{272}$110 
vanishes at $L \gt 22\hbar$. All isotopes of the element Z=110 have a rather
small fission barrier which disappears rapidly with increasing spin ($L$).
This is illustrated in Fig. 11, where the fission barriers of $^{266}$110
corresponding to a few $L$ are plotted as functions of the elongation.

In our calculations we only took into account those trajectories for
which the saddle point existed, i.e. we neglected the so--called quasifission
events. The number of emitted prescission particles is mostly governed by
the dynamics from the saddle to the scission point, but of course the initial 
conditions (see Section 3.1) play also an important role. In our model we
used two different sets of Langevin equations. The first set of equations
describes the fusion dynamics \cite{Fro90,Prz94} while the second one, 
coupled with the Master equations for particle emission, describes the 
fission process \cite{PBR96}. A better solution
would be a model in which the whole dynamics from fusion to fission
as well as the light particles emission would be described by one set of 
equations. Such an approach could immediately solve the problem of the proper 
setting of the initial conditions for the compound nucleus. 

The experimental and theoretical prefission neutron multiplicities for
$Z=$110 compound nuclei are written in Table 1 and additionally they are drawn 
in Fig.~12 as functions of the excitation  energy of the system.
The dashed lines correspond to the results obtained with the 
wall-and-window friction ($\gamma_{ww}$) reduced by 50\% while 
the dotted lines stand for the results obtained with the standard 
value of $\gamma_{ww}$. 
It is seen that for the lowest excitation energies the agreement of theoretical
results with the  measurements of neutron multiplicities is good,
while the discrepancy grows with increasing excitation energy. For the 
most excited compound nuclei our theoretical predictions are too large
by 1 to 2 units. This result could mean that the effective time which the
system needs to pass from the saddle to the scission point is too long 
in our model for a very hot compound nucleus. Any of transport coefficients 
do not depend on the temperature. This could be the reason of the observed 
discrepancy. In the near future we intend to perform a new calculation with 
temperature dependent transport coefficients evaluated within the linear 
response theory \cite{HI99,IH99}. The temperature dependence of the
friction parameter could be important, because as one can see in Fig.~12 
the reduction of the friction parameter by 50\% decrease the neutron 
multiplicities by 1 to 0.5 units depending on the excitation energy
of compound nucleus. 

In our model the superheavy nucleus with Z=110 is already cold when 
it reaches the scission point after emission of some neutrons
and $\alpha$-particles. So the shell effects begin to play an important
role for very deformed shapes. The shell effects could lead to the compact
fission, i.e. splitting of the Z=110 nucleus into two spherical fragments.
In our model the shell effects are not present but we simulate partially
their influence on the number prefission particles by counting only those 
particles which have been evaporated before reaching the elongation of nucleus 
close to the compact scission point ($R_{12}/R_0 \sim 1.5$).

Contrary to the case of decay of $^{126}$Ba and $^{188}$Pt compound nuclei
a hot superheavy nucleus with Z=110 exists and fissions at very low angular
momenta ($L \lt 20 \hbar$). This range of L corresponds to the linear
part of the differential fusion cross-section $d\sigma/dL$ (see Fig.~10),
so we do not expect entrance channel effects on multiplicities of
prefission neutrons.

The Coulomb barrier for emission of the charged particles from $Z=$110 isotopes
is rather high so the predicted proton and $\alpha$ multiplicities are
much smaller than the neutron multiplicities as one can see in Table~1.

\section{Summary, conclusions and further developments}

In the first part of the present work we described an extension of our
model \cite{PBR96} used so far for describing the fission dynamics of 
an excited, rotating and possibly deformed compound system. We considered
the decay of compound nuclei produced by the fusion of two heavy ions.
The initial spin distribution was determined by a simple model calculation
\cite{Fro90,Prz94} which took the possible deformation of the two ions into
account \cite{PPR94}. Furthermore, we introduced the microscopic classical 
expression for the emission rates of light particles derived in reference
\cite{DPR95}  and showed the link of this expression with
the Weisskopf formulation \cite{We37}.

In the second part of the paper we presented a detailed comparison 
of the calculated neutron multiplicities with the ones
obtained in different experiments involving systems from different regions
of the mass table. All five systems discussed in the paper ($^{126}$Ba,
$^{188}$Pt and Z=110 isotopes) are good examples of different fission
and particle evaporation mechanisms.

The light nucleus $^{126}$Ba fissions at very high angular momenta
($L \sim 80 \hbar$). Its saddle point is close to the scission point,
so the fission dynamics plays a rather minor role in this case. The
entrance channel effects influence significantly the decay properties
of the excited $^{126}$Ba, via different initial spin distributions
of compound nucleus. Also huge centrifugal forces and big deformations 
corresponding to the saddle point lead in this case to a true competition
between the neutron and $\alpha$-particle evaporation. 
The experimental prefission neutron multiplicities grow with the excitation
energy of $^{126}$Ba while it is not always the case in our theoretical 
estimates.

In the case of $^{188}$Pt the average angular momentum of the fissioning nucleus
is around $60 \hbar$, i.e. $20 \hbar$ less than for $^{126}$Ba. Also
the way from the saddle point to the scission point is longer so that the dynamics 
from the saddle to the scission plays here an important role. 
In addition, the evaporation of a few particles does not increase so 
dramatically the height of the fission barrier as in the case for 
$^{126}$Ba. As a consequence the fission process takes place during a longer 
time and the number of evaporated particles is larger.  Similarly 
as in the case of $^{126}$Ba the initial spin distribution (the entrance 
channel effect) influences the multiplicity of pre-fission neutrons
which is here too small (by $\approx$ 30\%) in comparison with the experimental
data.

A completely different case is the decay of superheavy compound nuclei
with Z=110. Here the way from the weakly deformed saddle point to the 
scission point is very long. The fission barrier of a hot compound nucleus
with Z=110 is very small and it vanishes already at low angular momenta
($L \lt 22 \hbar$).  The way in which the Z=110 nuclei has been produced
does not influence their decay properties since such nuclei
exist at small angular momenta only, which belong to the linear
part of the fusion cross-section $d\sigma/dL$.
A majority of compound nuclei goes to fission with simultaneous
emission of neutrons and $\alpha$-particles. The number of prefission
neutrons depends in this case on details of the fission dynamics, e.g.
the slope of the potential and the magnitude of the friction and 
diffusion parameters. 
Our model overestimates slightly ($\approx 15$\%) the experimental number
of prefission neutrons here.
The neutron emission process cools significantly
the Z=110 compound nuclei, so the temperature dependence of the
friction and diffusion parameters could be ``visible'' in this case.

The present formulation of the model and its application can be
improved on several points: 
\begin{itemize}
\item[(i)] One may introduce a more detailed description of the formation 
process of the compound nuclei. Such a description will
involve explicit trajectory calculations of the incident nuclei taking 
into account the nuclear and the Coulomb potentials for different relative 
initial orientations of the
deformed  nuclei.  Calculations of relative potentials have been worked
out in this framework \cite{PPR94,PPB99} but, to our knowledge, a full
fledged dynamical trajectory calculation has yet to be done.

\item[(ii)] The full treatment of fusion fission dynamics with simultaneous
emission of the light particles will release us from a somewhat arbitrary
choice of the initial conditions for the compound nucleus. Such
a project is at present under investigation.

\item[(iii)] We used a classical description of fluctuations and a form of
the fluctuation-dissipation theorem which are valid for the case of high 
temperature. In fact, the temperature of the systems which
we considered above could be low enough for quantum effects (pairing
and shell effects) to play a non negligible role. This point has been
investigated in the recent past \cite{HK98,HI99,IH99} within the framework 
of nuclear transport theory \cite{Hof97}. 
It is our aim to come back to this point in the near future.

\item[(iv)] Our fission barrier estimates need also further improvements.
The present model bases on a modified (see Appendix A) liquid drop formula 
of Myers and Swiatecki \cite{MS67} and it is known that for the cold light 
nuclei like $^{126}$Ba it overestimates the fission barriers by $\approx 10$ MeV 
\cite{Kra79}. 

\end{itemize}

For reasons of comparison with  experiment we shall 
generalize our model taking asymmetric fission
events into account. Many compound nuclei generated from mass asymmetric
formation process channels appear with a high statistical
weight. 

Last but not least it would be nice to extend the test of the model to 
observables other than the light particle
emission rates. Out-of plane particle angular distributions and
correlations between emitted light particles would provide further sensitive
tests of the validity of our approach. 
For instance, the ratio of neutrons emitted out of the reaction plane
and within the reaction plane (defined by the beam and the outgoing
fission fragments) would enable us to determine the deformation of
the emitting compound nucleus.

Furthermore, we stress the importance of a measurement of the emitted neutrons, 
protons, and $\alpha$-particles in the same system because the emission 
rates of these particles are reciprocally dependent on each other.

\bigskip\bigskip    
\noindent
{\bf Acknowledgements}
\bigskip

The authors J. B., K. D., and J. R. wish to
express their thanks for the warm hospitality extended to them at the
Institute for Theoretical Physics of the UMCS of Lublin.

\newpage
\begin{center}
{\bf Appendix A: Transport parameters used in the Langevin equation}
\end{center}

Fission dynamics correlated with prefission particle 
emission is generally described in a phenomenological framework, by means of 
quantities such as collective temperature-dependent potentials, masses,
moments of inertia and friction coefficients.

The fission process is described here in terms of the unique collective variable 
$q=R_{12}/R_0$. The fission path parameterized by $q$ is fixed by
means of a minimalization of the free energy of a nucleus in the
$(\alpha_0,\alpha_2,\alpha_4,\alpha_6)$ deformation space at a fixed 
temperature $T$. The deformation parameters are defined as follows
\cite{TKS80}
$$
\rho_s^2\,(z) = \rm{R}_0^2 \, \sum_{\ell=0}^6 \alpha_{2\ell}(q) \, 
P_{2\ell}\,(\frac{z}{z_0}) \;\; , \;\;\; -z_0 \leq z \leq z_0 \;\; ,
\eqno(A1)
$$
where the function $\rho_s\,(z)$ is the distance in cylindrical 
coordinates of any point of the surface to the symmetry axis.

The Helmholtz free energy of the nucleus plays the role of the collective 
potential and can be written, in a semiclassical approximation, as
$$
F\,(N,Z,q,L,T) = E(N,Z,q,T=0) \,- \, a(N,Z,q)\,T^2 \,+\, E_{rot}(N,Z,q,L) \;\;.
\eqno(A2)
$$
The first term on the r.h.s. of eq. (A2) is a liquid-drop (or any
macroscopic model) type energy expressing the semiclassical energy of
the deformed cold nuclear system as a function of the mass number 
$A \!= \! N \! + \! Z,$ and the isospin asymmetry $I =( N-Z)/A$. 
The deformation dependence of the free energy is taken
into account through the shape functions $B_s$ and $B_{coul}$. 
The rotational energy is calculated explicitly as
$E_{rot}=L^2/2{\cal J}(q)$ with a $q$-dependent rigid-body moment of
inertia. We have used in eq. (A2) the liquid drop parametrization of Myers 
and Swiatecki \cite{MS67}.
                                                                   
The effects of excitation are taken into account in eq. (A2) through the 
level-density parameter $a$ at a given deformation. As the Helmholtz free 
energy, which is the variational quantity, is given by 
$$
        F(T) = E(T) - T \, S(T)                                            
\eqno(A3)
$$
with the temperature T as a Langrange multiplier, one can show 
\cite{KR70,RKK72} that for a liquid-drop type system, entropy and excitation 
energy read
$$
          S = 2 \, a \, T                                                  
\eqno(A4)
$$
and
$$
          E^* = E(T) - E(T=0) = a \, T^2                                   
\eqno(A5)
$$
from which one obtains the last relation in eq. (A2). 

The level density parameter $a$ can be written in the form 
$$
a(N,Z,q) = a_v \, (1 + k_v I^2) \, A \, 
         + \, a_s \, (1 + k_s I^2) \, A^{2/3} B_s(q)  \nonumber\\
         + \, a_{coul} \, Z^2A^{-1/3} B_{coul}\,(q) \;\; ,
\eqno(A6)
$$
where the parameters $a_v$=0.0533 MeV$^{-1}$, $k_v$=0.5261, 
$a_s$=0.1059 MeV$^{-1}$, $k_s$=2.7192, and $a_{Coul}$=0.000458~MeV$^{-1}$
are taken from Ref. \cite{BGS88}.
In this expression, as well as in the macroscopic energy in eq. (A2), we have 
neglected the curvature and compression term proportional to $A^{1/3}$ since 
its coefficient for the cold system is known experimentally to be small 
\cite{MMS88}. When evaluating the Coulomb term in (A2) and (A6) the Coulomb 
exchange contribution is neglected. 

To describe the fission dynamics one also needs to 
calculate the mass parameter $M\,(q)$ which enters the
kinetic term of the equation of motion describing the fission process. In 
terms of the deformation parameter $q$ the mass is
given, in the incompressible fluid approximation \cite{Da76} , as
$$
   M= \pi \rho_0 \int\limits_{-z_0}^{z_0} [\rho_s^2(z) A^2(z)  +
                 {1 \over 2} B^2(z) ] dz \;\; ,
\eqno(A7)
$$
where
$$ A(z) = \frac{1}{\rho_s^2(z)} \frac{\partial}{\partial q} 
\int\limits_{z}^{z_0} \rho_s^2(z') dz' \;\; ,      
\eqno(A8)
$$
and 
$$ B(z) = {1 \over 2} \frac{\partial \rho_s^2}{\partial z} A  +  
\frac{\partial \rho_s^2}{\partial q}  
\eqno(A9)
$$
and where $\rho_0$=0.17 fm$^{-3}$ is the matter density.

The friction coefficient associated with the collective coordinate $q$ 
is calculated in the framework of the wall and window model 
\cite{Blo77,Blo78}. The wall contribution, which is the dominant part of 
the friction parameter $\gamma$ associated with the fission mode, is given 
by
$$
  \gamma = \frac{\pi}{2} \rho_0 \bar{v} \int\limits_{-z_0}^{z_0} 
               \frac{ \left(\frac{\partial\rho_s^2}{\partial q}\right)^2}
                    {\sqrt{\rho_s^2(z) + \frac{1}{4} 
               \left(\frac{\partial\rho_s^2}{\partial z}\right)^2}} \; dz\;\; .
\eqno(A10)
$$
where the average velocity $\bar{v}$ of the nucleons inside the nucleus is 
at zero temperature defined as 
$$
  \frac{\bar{v}}{c} = \frac{\bar{p}}{m c} 
                      = \frac{3}{4} \frac{\hbar}{m c} (3 \pi^2 \rho_0)^{1/3}
\eqno(A11)
$$
with the Fermi momentum $p_F = \hbar k_F = \hbar (3 \pi^2 \rho_0)^{1/3}$.

\newpage
\begin{center}
{\bf Appendix B: Comparison between the microscopic semi-classical
formulation of emission rates with the Weisskopf formula}
\end{center}

In Ref. \cite{DPR95}, a formulation of emission rates has been given which is
based on the picture of a Fermi gas of nucleons
at temperature $T$ being the confined to a deformed rotating
square well. In this appendix we shall show that in the special
case of a spherical well, the model of Ref. \cite{DPR95} becomes
equivalent to Weisskopf's emission rate formula \cite{We37,SDP91}.
We consider the example of neutron emission.

According to Ref. \cite{DPR95}, the total number $n$ of neutrons emitted
per time unit is given by
$$
 n = \int\limits_\Sigma d\sigma \int d^3p f_n(\vec r_0,\vec p)\,v_\perp(\vec r_0) \,
 w^{cl}_0(v_\perp(\vec r_0))\,,
\eqno{(B1)}
$$
where  $d\sigma$ is the infinitesimal element of the surface $\Sigma$
at the surface point $\vec r_0$. The momentum $\vec p$ and the velocity
$\vec v$ of a neutron are related by the equation
$$
 \vec p = m\vec v + m\vec\omega \times\vec r\,,
\eqno{(B2)}
$$
where $\vec r$ is the position vector, $m$ is the mass of the neutron,
and $\vec\omega$ is the angular velocity of rotational motion of the
nucleus. In the case of a spherical nucleus, a collective rotational
motion is not possible. Consequently, we may put $\vec\omega = 0$
without loss of generality.

The Wigner function $f_n(\vec r_0,\vec p)$ then describes a gas of
fermions confined to a spherical well of depth $V_0$ at a temperature
$T$. It is given by the expression
$$
 f_n (\vec r,\vec p) = {2\over h^3} \cdot {\theta(\vec r)\over 
1 + \exp\left[{1\over T}\left({p^2\over 2m} - V_0 -
\mu_n\right)\right]}\,, 
\eqno{(B3)}
$$
where $\theta(\vec r)$ is a step function defined to be 1 for position
vectors $\vec r$ of points inside the volume $\Omega$ enclosed by the
surface $\Sigma$ 
$$
\theta(\vec r) =
\left\{
\begin{array}{ll}
 1 & {\rm for}~\vec r\in\Omega~{\rm including}~\Sigma \\
 0 & {\rm otherwise}\,\,\,.
\end{array}\right.
\eqno{(B4)}
$$
The parameter $\mu_n$ is the chemical potential of the neutrons.

The last factor $w^{cl}_0$ in formula (B1) is the semi-classical
transmission coefficient for a neutron hitting the surface $\Sigma$ at
the point $\vec r_0$ with a velocity $v_\perp(\vec r_0)$ perpendicular
to the surface. For a spherical surface $\Sigma$ the unit vector
normal to the surface at the surface point $\vec r_0$ is given by the
radial unit vector 
$$
 \vec n_\Sigma(\vec r_0) = {\vec r_0\over R_0}\,,
\eqno{(B5)}
$$
where $R_0: = |\vec r_0|$ is the radius of the square well potential.

In Weisskopf's theory of evaporation, the total number of emitted
particles of a given sort is represented as an integral over the energy
$\varepsilon$ of the emitted particles \cite{We37,SDP91}. Consequently, 
we would like to rewrite the formula (B1) as an integral over 
the variable
$$
 \varepsilon = {p^2\over 2m} - V_0
\eqno{(B6)}
$$
performing at the same time the remaining integrations in a closed form. 
For this purpose let us first introduce polar coordinates ($p$,
$\theta_p$, $\varphi_p$) in momentum space choosing the surface unit vector
$\vec n_\Sigma(\vec r_0)$ as the polar axis
\begin{eqnarray}
&& p_\xi = p\sin\theta_p\cos\varphi_p\,,\nonumber \\
&& p_\eta = p\sin\theta_p\sin\varphi_p \,,\nonumber \\
&& p_\zeta = p\cos\theta_p\,. \nonumber
\end{eqnarray}
In these coordinates the perpendicular velocity $v_\perp(\vec r_0)$ is
given by
$$
 v_\perp(\vec r_0) = {p\cos\theta_p\over m}\,.
\eqno{(B7)}
$$
The transmission coefficient $w_0^{cl}$ can only be unequal to zero, if
the velocity $v_\perp(\vec r_0)$ is positive. 
Thus the polar angle $\theta_p$ can be restricted to the interval $0
\leq \theta_p \leq \pi/2$. The total neutron yield per time unit
is thus given by
$$
n = {4\pi\over h^3} \int\limits_\Sigma d\sigma \int\limits^{\infty}_0 dp
\int\limits^{{\pi\over 2}}_0 d\theta_p \cdot {p^3\cos\theta_p\sin\theta_p\over
m} \, w_0^{cl}\left({p\cos\theta_p\over m}\right) \cdot
\left[e^{{1\over T}\left({p^2\over 2m} - V_0 -
\mu_n\right)} + 1\right]^{-1}\,.
\eqno{(B8)}
$$
The surface integral yields for factor $4\pi R^2_0$. Instead of
integration variables $p$ and $\theta_p$, we introduce the energy
$\varepsilon$ and the orbital angular momentum $l$ of the neutron
impinging on the surface
$$
 \varepsilon = {p^2\over 2m} - V_0 \,,
\eqno{(B9)}
$$
$$
 l = R_0\, p\sin\theta_p\,.
\eqno{(B10)}
$$
The Jacobian $D$ of the transformation
$$
 D = \left|
\begin{array}{ll}
{\partial p\over \partial\varepsilon} & {\partial \theta_p\over
\partial\varepsilon}\\ 
 & \\
{\partial p \over \partial l} & {\partial \theta_p \over \partial l}
 
\end{array}\right|
\eqno{(B11)}
$$               
 is obtained from the inverse transformation
$$
\left\{
\begin{array}{l}
  p =\sqrt{2m(\varepsilon + V_0)} \nonumber\\
  \theta_p = \arcsin \left({l\over R_0p}\right) = 
 \arcsin\left({l\over R_0\sqrt{2m (\varepsilon + V_0)}}\right)
\end{array}
\right.
\eqno{(B12)}
$$
in the form
$$
 D  
		= { \sqrt{2m/(\varepsilon+V_0)}
		\over 2\sqrt{2mR^2_0(\varepsilon+V_0)-l^2}} \,.
\eqno{(B13)}
$$ 
In formulating the integration limits for the new variables $\varepsilon$ and $l$
we must take into account the condition
$$
 {l^2\over 2mR^2_0} \leq {p^2\over 2m} = \varepsilon + V_0\,,
\eqno{(B14)}
$$
which implies that the total kinetic energy of the neutron can not be smaller
than its rotational part.

The factor $p^3\cos\theta_p\sin\theta_p$ can be written in terms of the new 
variables as follows
$$
 p^3\cos\theta_p\sin\theta_p = 2m(\varepsilon + V_0) \sqrt{1 - {l^2\over 
 2mR^2_0(\varepsilon + V_0)}} \cdot {l\over R_0}\,.
\eqno{(B15)}
$$
Making use the Eqs. (B13) to (B15) we obtain for the r.h.s. 
of Eq. (B8):
\medskip
$$
\begin{array}{ccc}
 n&=&2\pi \cdot {2\over h^3} \,4\pi R^2_0\int\limits^{\infty}_{-V_0}
  d\varepsilon
  \int\limits^{\sqrt{2mR^2_0(\varepsilon+V_0)}}_0 dl \,{1\over 2}
	\sqrt{2m\over (\varepsilon+V_0)} \cdot   \\
 & & \\
&\cdot& {2(\varepsilon + V_0)\over\sqrt{2mR^2_0(\varepsilon+V_0) - l^2}} \cdot 
  \sqrt{1 - {l^2\over 2mR^2_0(\varepsilon + V_0)}}\cdot {l\over R_0}
  \cdot {w_0^{cl}(\varepsilon, l)\cdot
	 \over \left[e^{(\varepsilon-\mu_n)/T} + 1 \right]}\,,
\end{array}
\eqno{(B16)}
$$
\medskip
or
$$
 n = {(4\pi)^2\over h^3} \int\limits^{\infty}_{-V_0} d\varepsilon \cdot 
  \int\limits^{\sqrt{2mR^2_0(\varepsilon+V_0)}}_0 dl \cdot l \cdot 
  {w_0^{cl}(\varepsilon, l)
	 \over \left[e^{(\varepsilon-\mu_n)/T} + 1\right]}\,.
\eqno{(B17)}
$$

In Eq. (B17), the transmission coefficient $w_0^{cl}(\varepsilon,l)$ 
is obtained from 
$w_0^{cl}(v_\perp)$ by expressing the argument $v_\perp$ by the variables 
$\varepsilon$ and $l$.
$$
 v_\perp(\varepsilon,l) =  {\sqrt{2mR^2_0(\varepsilon + V_0) 
	 - l^2}\over mR_0}\,.
\eqno{(B18)}
$$
Introducing the dimensionless angular momentum $\lambda$ instead of the 
variable $l$
$$
 l = \hbar\lambda\,,
\eqno{(B19)}
$$
 the formula (B17) takes the form
$$
n = \int\limits^\infty_{-V_0} d\varepsilon 
\int\limits^{\lambda_{\rm max}(\varepsilon)}_0
 d\lambda \, {\partial^2 n\over\partial\varepsilon\partial\lambda} \,,
\eqno{(B20)}
$$
with
$$
 {\partial^2 n\over\partial\varepsilon\partial\lambda}  = {4\over h}
 {\lambda \cdot w^{cl}(\varepsilon,\lambda)\over \left[e^{
 (\varepsilon-\mu_n)/T} + 1 \right]} 
\eqno{(B21)}
$$
and
$$
 \lambda_{\rm max}(\varepsilon): = {1\over\hbar}\sqrt{2mR^2_0(\varepsilon
  + V_0)}\,.
\eqno{(B22)}
$$
We wish to compare the result (B21) with the partial width 
$\Gamma^{\alpha\beta}_\nu(E^*,\Lambda)$ for the decay of a nucleus 
of excitation energy $E^*$ and vanishing
total angular momentum $(\Lambda = 0)$ by emission of a neutron of energy
$\varepsilon_\alpha$ and orbital angular momentum $\lambda_\beta$ 
as it is obtained from Weisskopf's theory (Refs. \cite{PBR96}, \cite{DPR95})
$$
\Gamma^{\alpha\beta}_n = {2(2\lambda_\beta + 1)
\over h\,\rho(E^*,\Lambda=0)}
 \cdot \rho_R(E^*_R,\Lambda_R) \, w_n(\varepsilon_\alpha, 
 \lambda_\beta)\,.
\eqno{(B23)}
$$
In (B23), $\rho(E^*,\Lambda)$ and $\rho_R(E^*_R,\Lambda_R)$ are the level
densities for the mother and daughter nucleus, resp., which depend on
the excitation energy $E^*(E^*_R)$ and the total angular momentum $\Lambda
(\Lambda_R)$ of the mother (daughter) nucleus. The quantity 
$w_n(\varepsilon_\alpha,
\lambda_\beta)$ is the transmission factor for a neutron having the energy
$\varepsilon_\alpha$ and the orbital angular momentum $\lambda_\beta$.
The factor 2 $(2\lambda_\beta + 1)$, represents the product of the degeneracy
factors of the emitted neutron and of the residual daughter nucleus.
We note that, in the derivation of Eq. (B23) and also in our model, 
effects of the spin-orbit coupling on the emission probability are neglected.
We now use the level densities obtained in the Fermi gas model:
$$
 \rho_R(E^*_R, 0) = \left({\hbar^2\over 2J}\right)^{3/2}
 \sqrt{a}\, {\exp(2\sqrt{aE^*_R}) \over 12E^{*2}_R} \,,
\eqno{(B24)}
$$
$$
 \rho(E^*,0) = \left({\hbar^2\over 2J}\right)^{3/2}
 \sqrt{a} \, {\exp(2\sqrt{aE^*}) \over 12E^{*2}} \,.
\eqno{(B25)}
$$
Here $J$ is the rigid body moment of inertia of the nucleus which is assumed 
to be the same for the mother and daughter nucleus, and ''$a$'' is the level 
density parameter.

The excitation energies $E^*$ and $E^*_R$ are related by the energy conservation
$$
 E^* = E^*_R + \varepsilon_\alpha - \mu_n\,.
\eqno{(B26)}
$$
The ratio of the level densities is thus given by
$$
\begin{array}{l}
 {\rho_R\over\rho} = \left(\exp\left[2\sqrt{aE^*_R} - 2\sqrt{aE^*}\right]
\right) \cdot \left({E^*\over E^*_R}\right)^2 \\ 
\\
{\rho_R\over\rho} = \left(\exp\left[2\sqrt{aE^*} \cdot \left(\sqrt{1 -
{\varepsilon_\alpha - \mu_n\over E^*}} - 1\right)\right]\right) \cdot \left(1
- {\varepsilon_\alpha - \mu_n\over E^*}\right)^{-2}\,. 
\end{array}
\eqno{(B27)}
$$
Assuming that the ratio $(\varepsilon_\alpha - \mu_n)/E^*$ is much smaller than 1
$$
 {\varepsilon_\alpha - \mu_n\over E^*} << 1
\eqno{(B28)}
$$
and using the relation between the excitation energy $E^*$ and the temperature 
$T$ as given by the Fermi gas model
$$
 E^* = aT^2
\eqno{(B29)}
$$
we obtain for the ratio of the level densities
$$ 
{\rho(E^*_R)\over \rho(E^*)} \approx \exp\left(-{\varepsilon_\alpha - 
\mu_n\over T}\right)\,.
\eqno{(B30)}
$$
Using (B30) we obtain for the neutron width (B23)
$$
 \Gamma^{\alpha\beta}_n \approx {2(2\lambda_\beta + 1)\over h} \,
 e^{-{\varepsilon_\alpha - \mu_n\over T}} \cdot w_n (\varepsilon_\alpha, 
 \lambda_\beta)\,.
\eqno{(B31)}
$$
This form of Weisskopf's general formula agrees indeed with our result (B21)
 if the transmission factor $w_n$ is calculated classically ($w_n = w_0^{cl}$), 
 if the orbital angular momentum $\lambda_\beta$ is large enough so that
$$
 2\lambda_\beta + 1 \approx 2\lambda_\beta
\eqno{(B32)}
$$
and if
$$
 \exp\left({\varepsilon_\alpha - \mu_n\over T}\right) >> 1\,,
\eqno{(B33)}
$$
so that the Fermi occupation factor in (B21) becomes equivalent to 
the Boltzmann factor
$$ 
{1 \over e^{\varepsilon_\alpha-\mu_n\over T}+1} \approx
e^{-{\varepsilon_\alpha-\mu_n\over T}}\,. 
\eqno{(B34)}
$$
We note that the approximation (B34) is already used for obtaining the
level density formula (B25) from the Fermi-gas model. 

We would like to comment that the well-known formula (B23) is already a
slight extension of the original result of Weisskopf published in Ref.
\cite{We37}. There, the transmission factor $w_n$ is expressed in terms of the
cross-section for the absorption of a neutron by the daughter nucleus
using detailed balance. In this form, Weisskopf's result is of very
general validity. The difficulty with it is only that detailed balance
relates the transmission factor $w_n$ to the cross-section for
absorption of a neutron by a nucleus with excitation energy $E^*_R$ and
angular momentum $\Lambda_R$. This absorption cross-section can, of
course, not be measured. Therefore, one needs a model to calculate the
transmission factor. 

The considerations in this appendix demonstrate at the same time that
it is meaningful to replace the purely classical transmission factor
$w_0^{cl}$ 
$$ 
 w_0^{cl}(\varepsilon,l) = \theta_0\left(\varepsilon + V_0 - {l^2 \over
 2MR^2_0}\right) 
\eqno{(B35)}
$$
by a quantum-mechanical barrier penetration factor as it is done in all
our numerical calculations.

\newpage

{\bf\large Table captions:}
\begin{enumerate}
\item Multiplicities of the prefission particles emitted by:  $^{126}$Ba 
 $^{188}$Pt and $^{266,272,278}$110 at different excitation energies. 
 The theoretical
 estimates for the compound nuclei with Z=110 are done for two values of
 the friction force. In the first rows are the data evaluated with the
 friction reduced by 50\% while those in the second row correspond to
 the standard wall-and-window friction.
\end{enumerate}

\vspace{2cm}
{\bf\large Figure captions:}
\begin{enumerate}
\item The differential fusion (solid lines) and fission (bars) cross 
       sections  for the compound nucleus $^{126}$Ba for different
entrance 
       channel reactions.

\item The deformation potential $V_{\rm fiss}$ for $^{126}$Ba as a function
       of the ''fission variable'' ${R_{12}/R_0}$ from the ground state
       deformation up to the scission point. The different curves correspond to
       the different values of the angular momentum $L$. The temperature of 
       $^{126}$Ba is fixed at $T$=1.6 MeV.
       
\item The same as in Fig. 2, but now the different curves correspond to the
       different temperatures $T$. All plots are done for $L$=30$\hbar$.
       The different curves are shifted vertically in order to
       make the relative changes more visible. 

\item The temperature of the compound nucleus $^{126}$Ba at the different
       excitation energies as a function of angular momentum.

\item The multiplicities of prefission particles as a function of excitation
       energy of the compound nucleus $^{126}$Ba obtained with the
entrance
       channel $^{28}$Si + $^{98}$Mo.
       
\item The same as in Fig.~5 but for the reaction $^{19}$F + $^{107}$Ag.       

\item The same as in Fig. 1 but for the compound nucleus $^{188}$Pt.

\item The same as in Fig. 2 but for the compound nucleus $^{188}$Pt.

\item The same as in Fig. 3 but for the compound nucleus $^{188}$Pt.
 
\item The same as in Fig. 1 but for the compound nucleus $^{272}$110.

\item The same as in Fig. 2 but for the compound nucleus $^{266}$110
       and temperature $T$=0.

\item The experimental (points with error bars) and theoretical 
       multiplicities of prefission neutrons as a function of the excitation
       energy of the compound nuclei with $Z=$110 obtained in four different
       reactions.

\end{enumerate}

\newpage

\begin{center} {\large\bf Table 1} \end{center}

\begin{center}
\begin{tabular}{|c|c|c|c|c|c|c|c|c|c|c|}
\hline
CN&Reaction & $E_{lab}$ &$E^*$&$M_n^{exp}$&$\delta M_n^{exp}$&
                                             $M_n$&$M_p$&$M_\alpha$\\
\hline
 & &MeV&MeV&-&-&-&-&-\\
\hline
&                      & 204.0 & 131.7 & 2.52 & 0.12 & 2.29 & 0.03 & 0.79
\\
&$^{28}$Si + $^{98}$Mo & 
                         187.2 & 118.5 & 2.01 & 0.13 & 1.71 & 0.00 & 0.09
\\
$^{126}$Ba&            & 165.8 & 101.4 & 1.32 & 0.09 & 1.83 & 0.00 & 0.04
\\
&                      & 142.8 &  84.1 &  -   &  -   & 0.27 & 0.04 & 0.88
\\
\cline{2-9}
&$^{19}$F + $^{107}$Ag & 
                         147.8 & 118.5 & 1.85 & 0.11 & 1.99 & 0.00 & 0.16
\\
&                      & 128.0 & 101.5 & 1.31 & 0.17 & 1.80 & 0.01 & 0.06
\\
\hline
&$^{34}$S + $^{154}$Sm & 202.6 & 100.0 & 4.5  & 0.7  & 3.52 & 0.00 & 0.10
\\
$^{188}$Pt &           & 159.8 &  66.5 & 2.5  & 0.7  & 1.10 & 0.00 & 0.00
\\
\cline{2-9}    
&$^{16}$O + $^{172}$Yb & 137.6 &  99.7 & 5.4  & 0.6  & 3.79 & 0.00 & 0.06
\\
\hline
&                      & 513.9 & 185.9 & 7.83 & 0.46 & 8.66 & 0.26 & 1.04 \\
&                      &       &       &       &     & 9.72 & 0.23 & 0.95 
\\ 
&                      & 486.6 & 164.6 & 7.35 & 0.50 & 7.65 & 0.20 & 0.93 \\
&                      &       &       &       &     & 8.56 & 0.21 & 0.79 
\\ 
$^{266}$110&
 $^{58}$Ni + $^{208}$Pb& 461.1 & 144.6 & 6.17 & 0.48 & 6.50 & 0.16 & 0.81 \\
&                      &       &       &       &     & 7.30 & 0.17 & 0.73 
\\ 
&                      & 436.2 & 125.1 & 4.74 & 0.49 & 5.45 & 0.10 & 0.63 \\
&                      &       &       &       &     & 6.21 & 0.11 & 0.58 
\\ 
&                      & 410.1 & 104.7 & 4.41 & 0.41 & 4.35 & 0.07 & 0.48 \\
&                      &       &       &       &     & 4.94 & 0.07 & 0.48 
\\ 
&                      & 377.0 &  78.8 & 2.94 & 0.36 & 2.77 & 0.03 & 0.30 \\
&                      &       &       &       &     & 3.23 & 0.03 & 0.33 
\\ 
\hline
&                      & 472.3 & 138.3 & 5.98 & 0.43 & 7.49 & 0.08 & 0.42 \\
&                      &       &       &      &      & 8.25 & 0.08 & 0.34 
\\
&                      & 449.9 & 121.1 & 5.52 & 0.38 & 6.40 & 0.05 & 0.30 \\
&                      &       &       &      &      & 7.12 & 0.06 & 0.28
\\
&$^{64}$Ni + $^{208}$Pb& 423.0 & 100.6 & 5.13 & 0.33 & 5.09 & 0.02 & 0.20 \\
&                      &       &       &      &      & 5.58 & 0.04 & 0.20
\\
&                      & 403.9 &  85.9 & 3.54 & 0.31 & 3.94 & 0.01 & 0.15 \\
&                      &       &       &      &      & 4.49 & 0.01 & 0.12
\\
$^{272}$110&
                       & 377.0 &  65.3 & 3.03 & 0.32 & 2.27 & 0.00 &  0.10 \\
&                      &       &       &      &      & 2.83 & 0.00 &  0.08
\\
\cline{2-9}
&                      & 351.2 & 166.3 & 8.40 & 0.53 & 9.23 & 0.13 &  0.53 \\
&                      &       &       &      &      & 9.95 & 0.15 &  0.48
\\
&$^{40}$Ca + $^{232}$Th& 298.4 & 121.3 & 5.81 & 0.50 & 6.40 & 0.05 &  0.30 \\
&                      &       &       &      &      & 7.12 & 0.06 & 0.28
\\
&                      & 250.4 &  80.3 & 3.35 & 0.34 & 3.56 & 0.01 &  0.14 \\
&                      &       &       &      &      & 4.04 & 0.01 & 0.12
\\
\hline
&                      & 299.6 & 127.2 & 5.78 & 0.51 & 7.89  & 0.03 & 0.14 \\
&                      &       &       &      &      & 8.48 & 0.04 & 0.15
\\
$^{278}$110&
$^{40}$Ar + $^{238}$U  & 280.4 & 110.7 & 4.96 & 0.55 & 6.69 & 0.00 &  0.12 \\
&                      &       &       &      &      & 7.23 & 0.03 & 0.13
\\
&                      & 258.0 &  91.5 & 4.22 & 0.44 & 5.08 & 0.00 & 0.09 \\
&                      &       &       &      &      & 5.59 & 0.00 & 0.09
\\
\hline
\end{tabular}
\end{center}

\pagestyle{empty}

\newpage
\begin{figure}
\epsfysize=180mm  \epsfbox{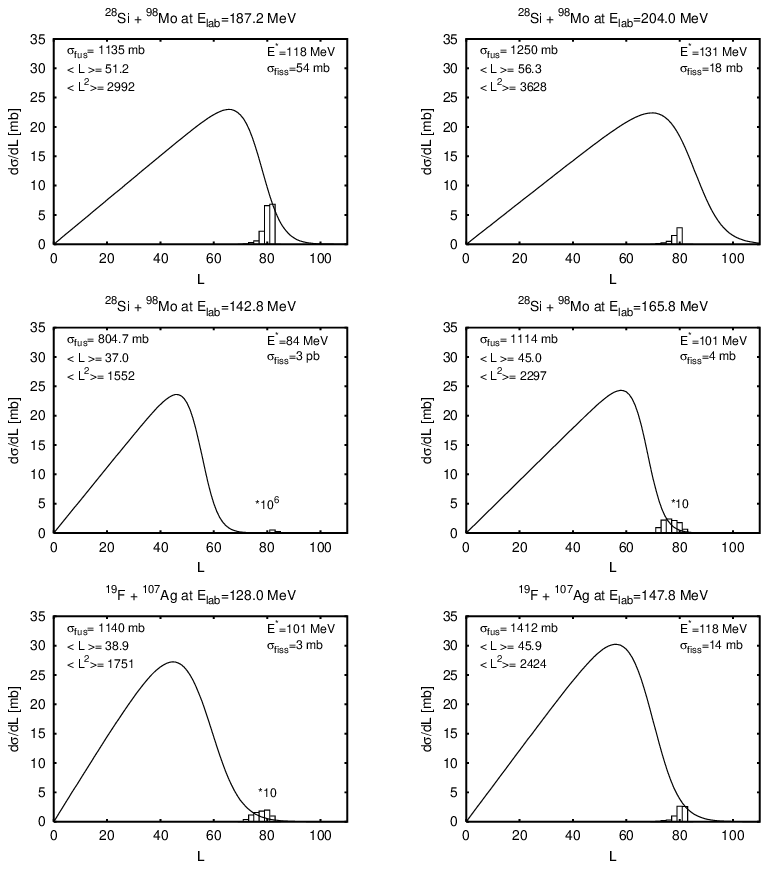}
\caption{}
\end{figure}

\newpage
\begin{figure}
\epsfxsize=140mm  \epsfbox{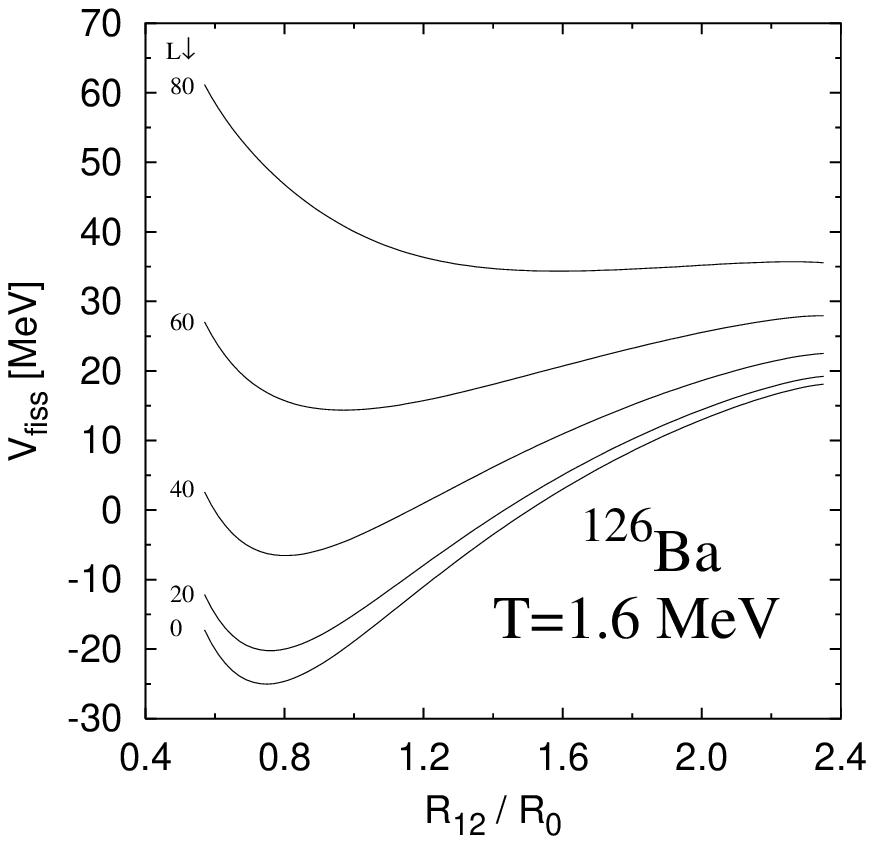}
\caption{}
\end{figure} 

\newpage
\begin{figure}
\epsfxsize=140mm  \epsfbox{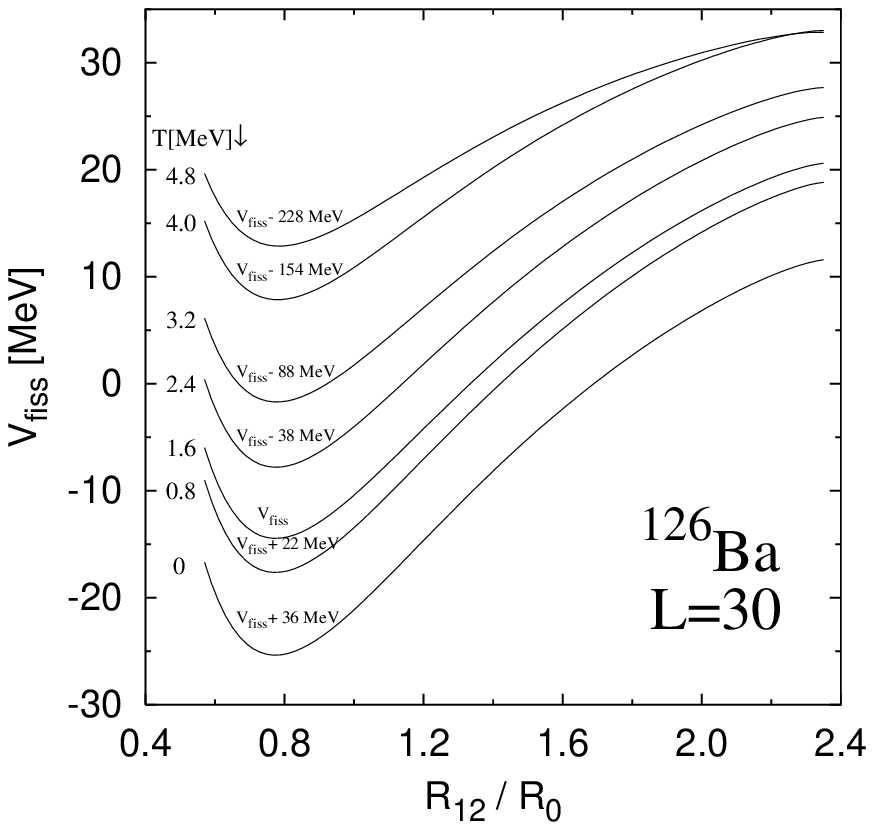}
\caption{}
\end{figure} 

\newpage
\begin{figure}
\epsfxsize=140mm  \epsfbox{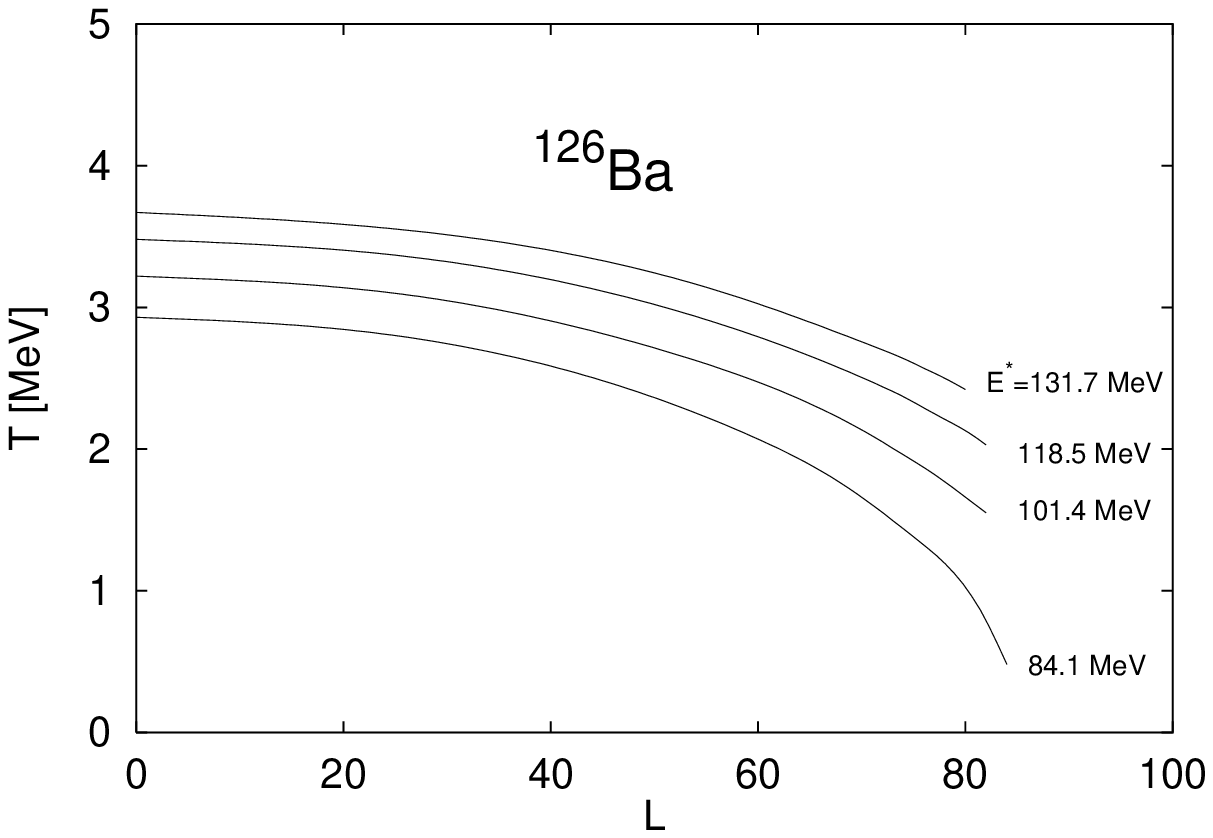}
\caption{}
\end{figure} 

\newpage
\begin{figure}
\epsfxsize=140mm  \epsfbox{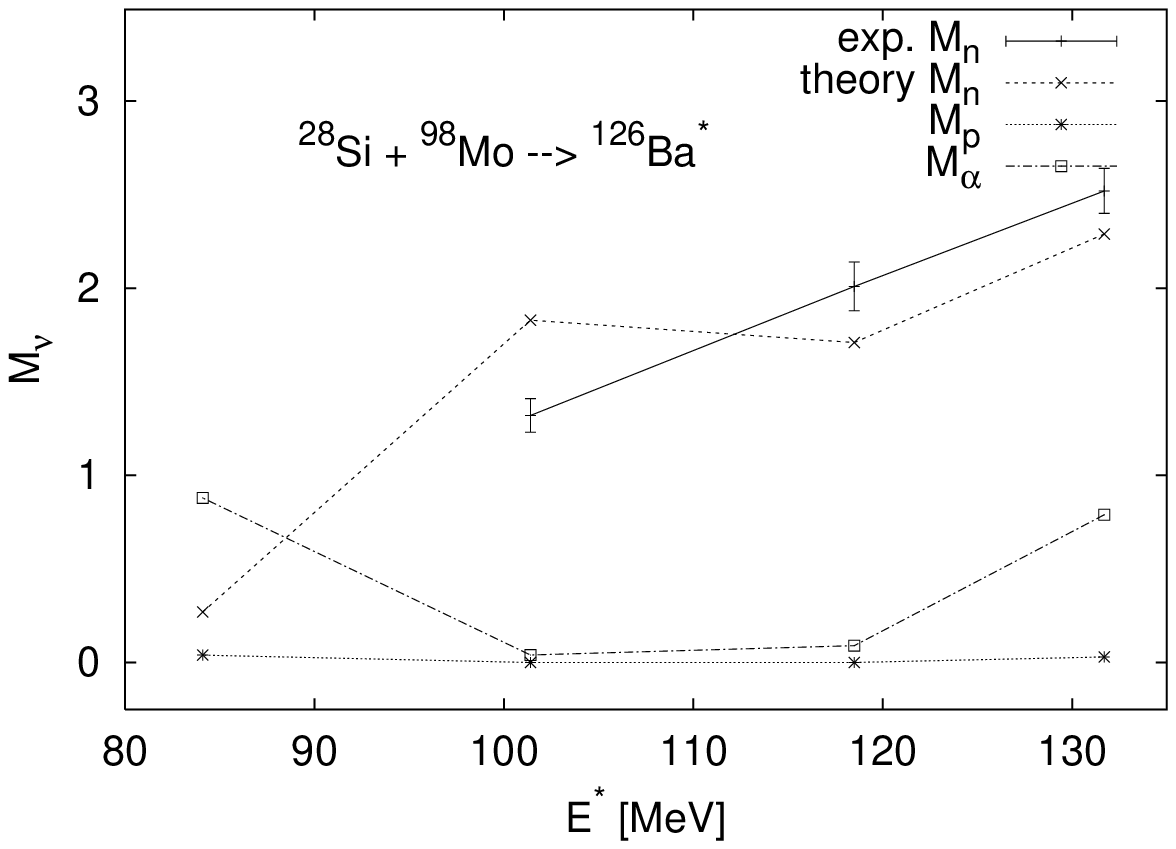}
\caption{}
\end{figure} 

\newpage
\begin{figure}
\epsfxsize=140mm  \epsfbox{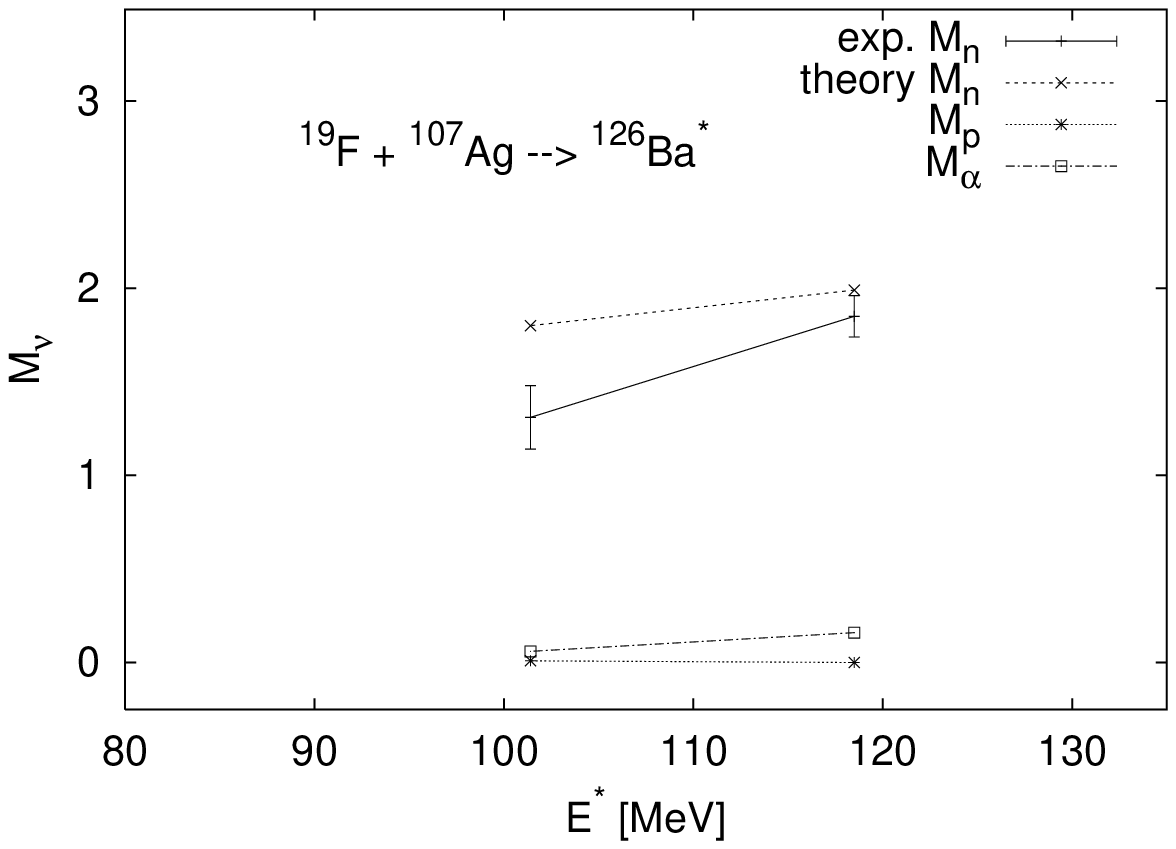}
\caption{}
\end{figure} 

\newpage
\begin{figure}
\epsfxsize=170mm  \epsfbox{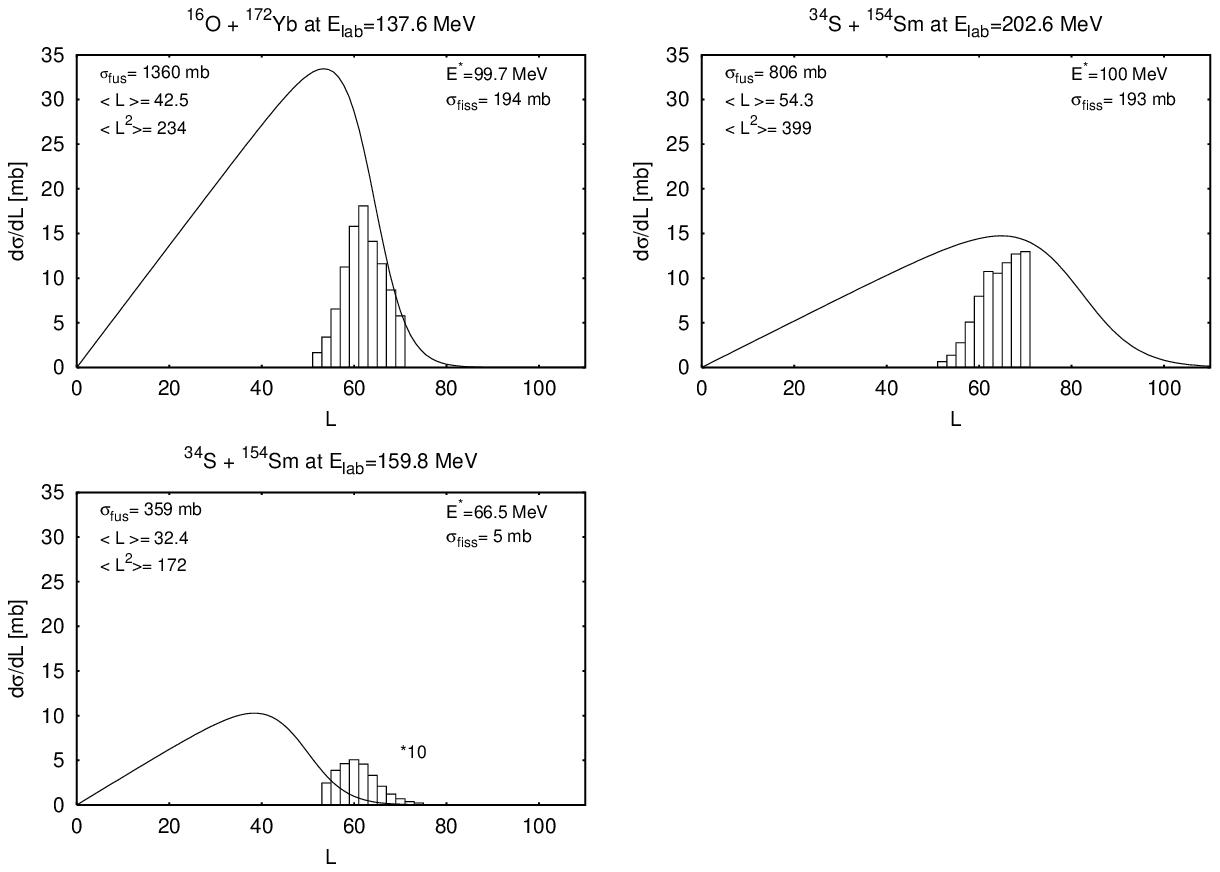}
\caption{}
\end{figure} 

\newpage
\begin{figure}
\epsfxsize=140mm  \epsfbox{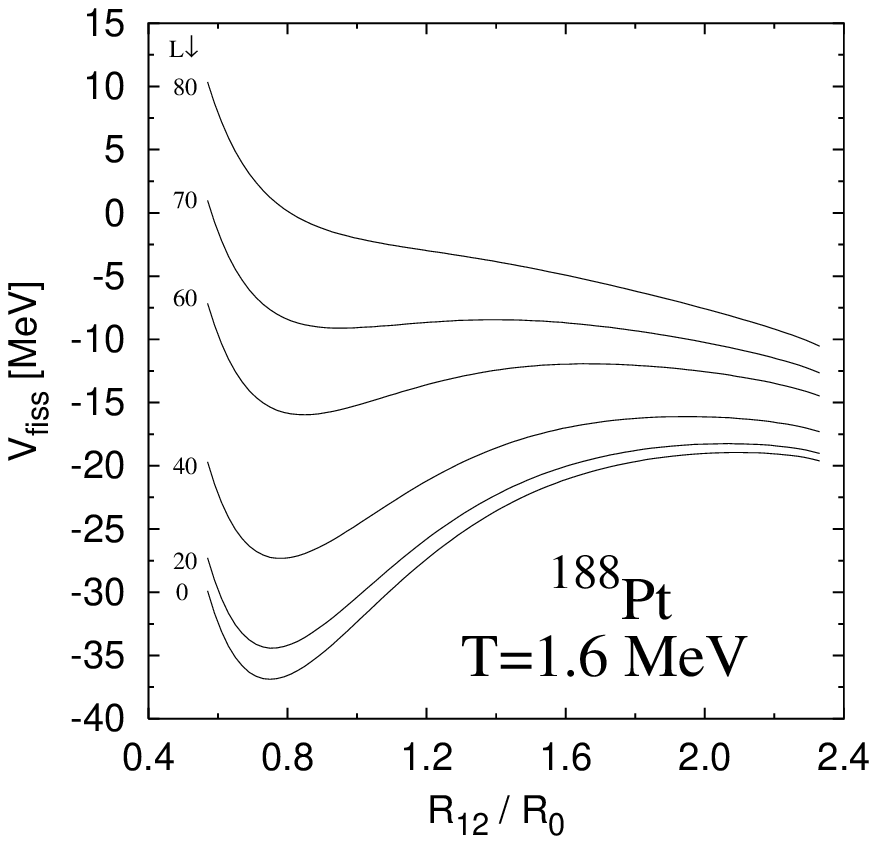}
\caption{}
\end{figure} 

\newpage
\begin{figure}
\epsfxsize=140mm  \epsfbox{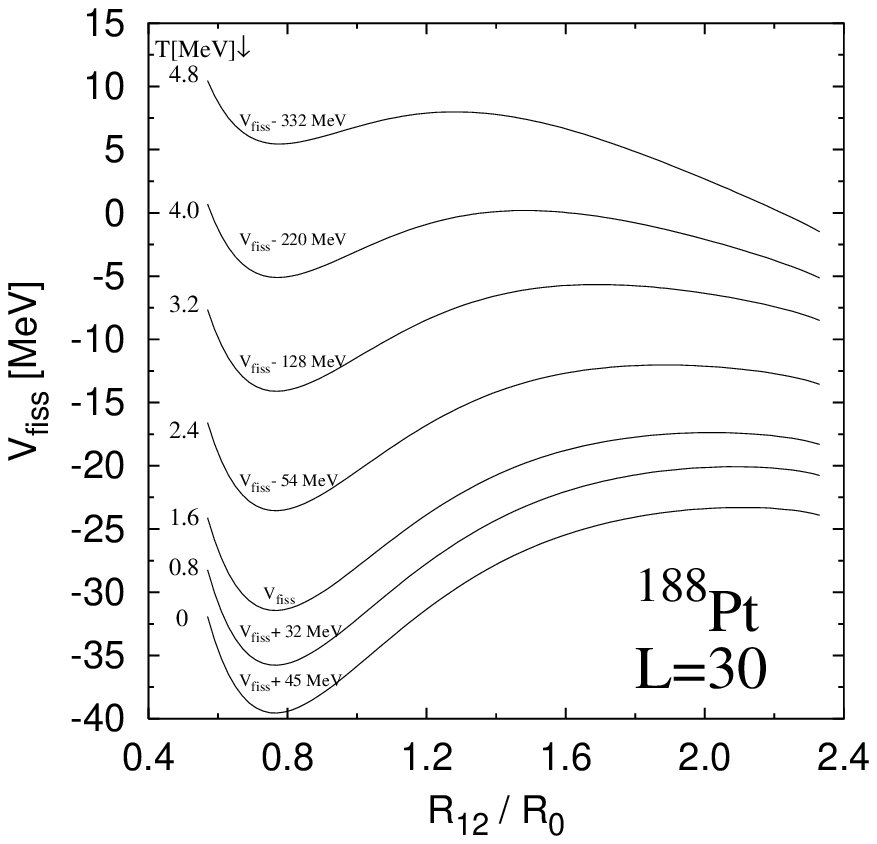}
\caption{}
\end{figure} 

\newpage
\begin{figure}
\epsfxsize=140mm  \epsfbox{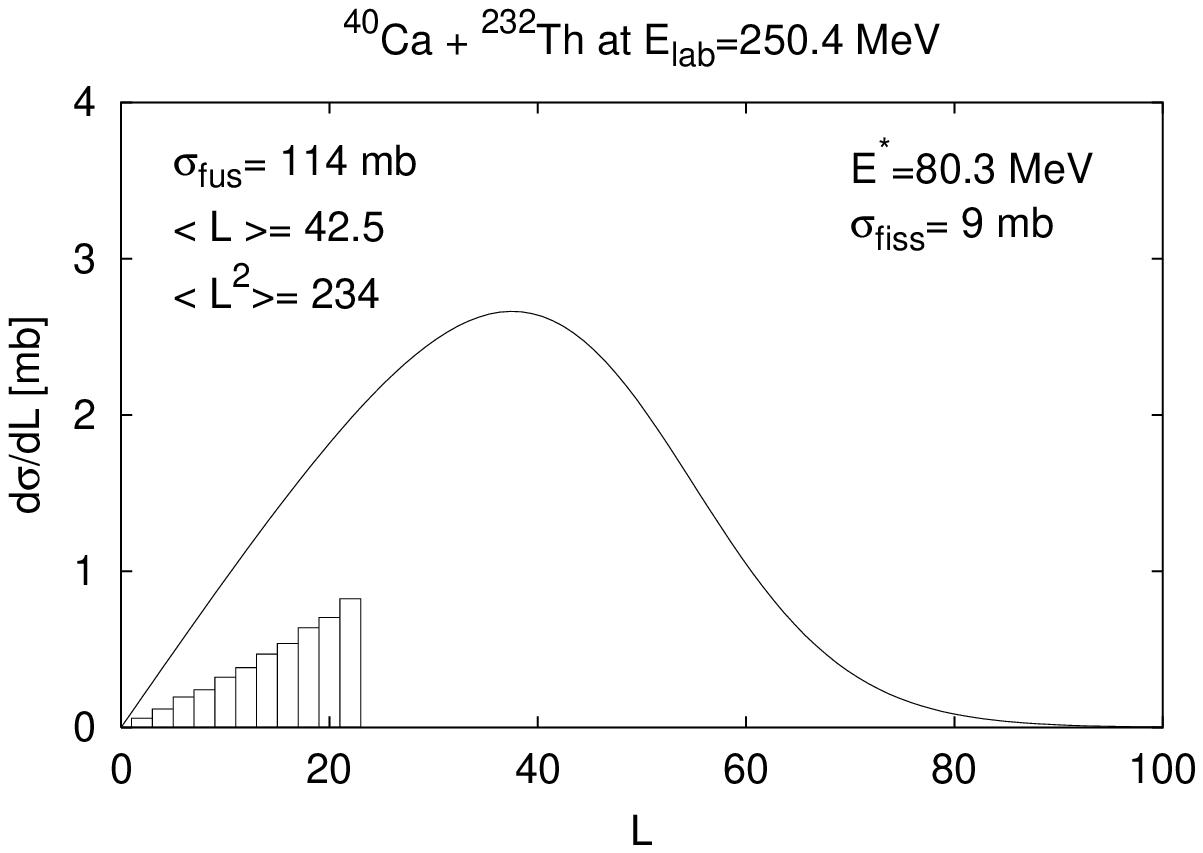}
\caption{}
\end{figure} 

\newpage
\begin{figure}
\epsfxsize=140mm  \epsfbox{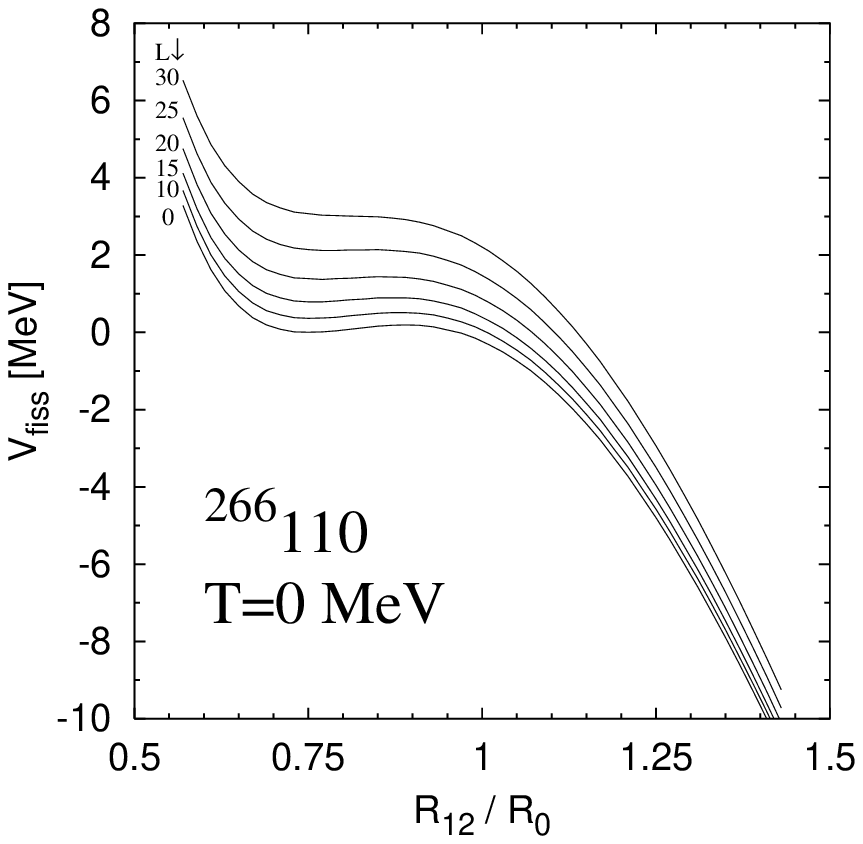}
\caption{}
\end{figure}

\newpage
\begin{figure}
\epsfxsize=165mm  \epsfbox{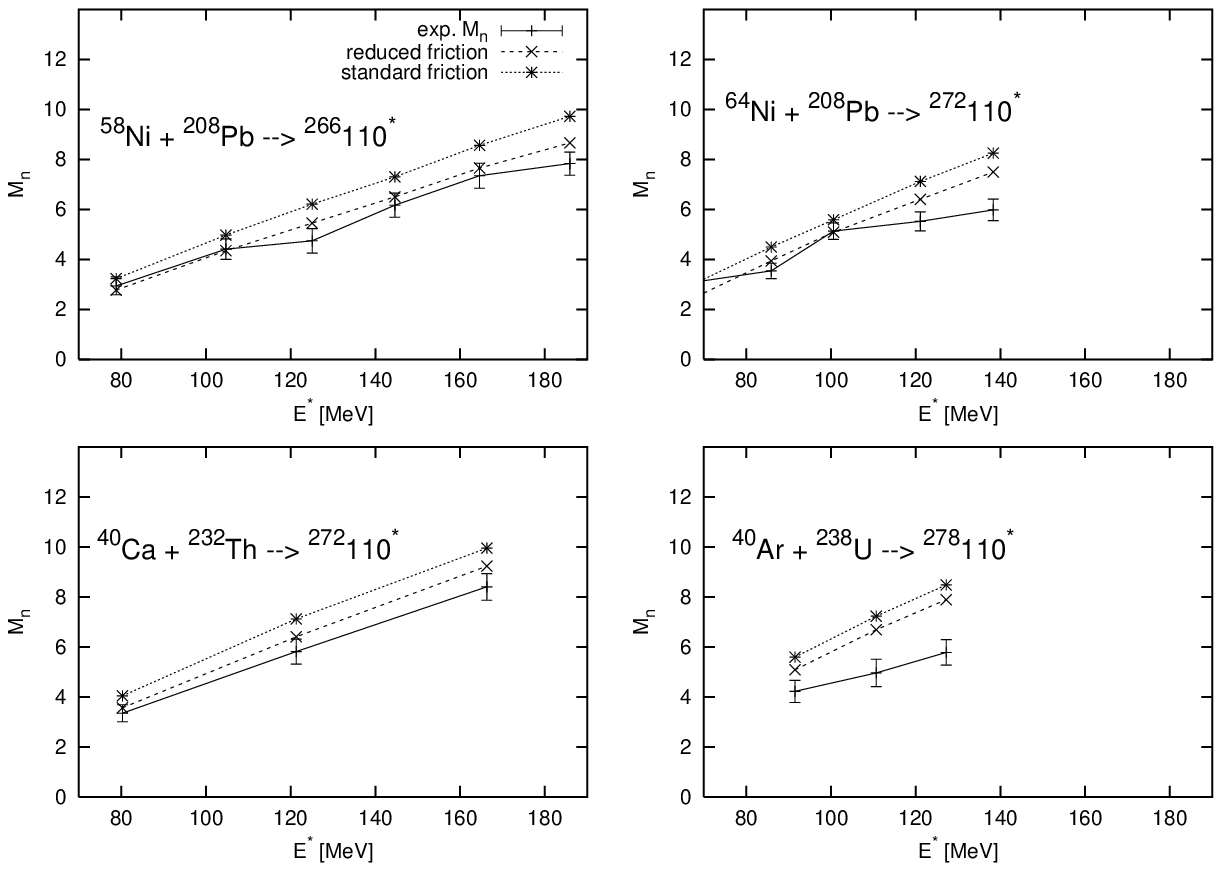}
\caption{}
\end{figure} 

\end{document}